\documentclass[aps,preprint,amsmath,amssymb,superscriptaddress]{revtex4-2}
\usepackage{times}
\usepackage{graphicx}
\usepackage{dcolumn}
\usepackage{bm}
\usepackage{comment}
\usepackage{lipsum}
\usepackage{hyperref}
\usepackage{color}
\usepackage{epsfig}
\usepackage{latexsym}
\usepackage{mathtools}
\usepackage{caption}
\usepackage{subcaption}
\usepackage{bbold}
\usepackage{ulem}
\usepackage{verbatim}
\usepackage{url}
\usepackage{amsmath}
\usepackage[version=4]{mhchem}
\begin{document}

\title{Emergent Disorder and Mechanical Memory in Periodic Metamaterials}

\author{Chaviva Sirote-Katz}
\thanks{These authors contributed equally to this work}
\affiliation{Department of Biomedical Engineering, Tel Aviv University, Tel Aviv 69978, Israel}
\author{Dor Shohat}
\thanks{These authors contributed equally to this work}
\affiliation{School of Physics and Astronomy, Tel Aviv University, Tel Aviv 69978, Israel}
\author{Carl Merrigan}
\affiliation{School of Mechanical Engineering, Tel Aviv University, Tel Aviv 69978, Israel}
\author{Yoav Lahini}
\affiliation{School of Physics and Astronomy, Tel Aviv University, Tel Aviv 69978, Israel}
\affiliation{Center for Physics and Chemistry of Living Systems, Tel Aviv University, Tel Aviv 69978, Israel}
\author{Cristiano Nisoli}
\affiliation{Theoretical Division, Los Alamos National Laboratory, Los Alamos NM 87545, USA}
\author{Yair Shokef}
\email{shokef@tau.ac.il}
\affiliation{School of Mechanical Engineering, Tel Aviv University, Tel Aviv 69978, Israel}
\affiliation{Center for Physics and Chemistry of Living Systems, Tel Aviv University, Tel Aviv 69978, Israel}
\affiliation{Center for Computational Molecular and Materials Science, Tel Aviv University, Tel Aviv 69978, Israel}
\affiliation{International Institute for Sustainability with Knotted Chiral Meta Matter, Hiroshima University, Japan}

\begin{abstract}
Ordered mechanical systems typically have one or only a few stable rest configurations, and hence are not considered useful for encoding memory. Multistable and history-dependent responses usually emerge from quenched disorder, for example in amorphous solids or crumpled sheets. In contrast, due to geometric frustration, periodic magnetic systems can create their own disorder and espouse an extensive manifold of quasi-degenerate configurations. Inspired by the topological structure of frustrated artificial spin ices, we introduce an approach to design ordered, periodic mechanical metamaterials that exhibit an extensive set of spatially disordered states. While our design exploits the correspondence between frustration in magnetism and incompatibility in meta-mechanics, our mechanical systems encompass continuous degrees of freedom, and are hence richer than their magnetic counterparts. We show how such systems exhibit non-Abelian and history-dependent responses, as their state can depend on the order in which external manipulations were applied. We demonstrate how this richness of the dynamics enables to recognize, from a static measurement of the final state, the sequence of operations that an extended system underwent. Thus, multistability and potential to perform computation emerge from geometric frustration in ordered mechanical lattices that create their own disorder.
\end{abstract}

\maketitle

\newpage

\section{Introduction}

When interactions between different components of a complex system cannot simultaneously be minimized~\cite{ramirez1994strongly, moessner2006geometrical}, the lowest-energy state is a compromise in which some elements of the system are left ``unhappy". This frustration can lead to constrained disorder, degeneracy, and multistability~\cite{wannier1950antiferromagnetism, harris1997geometrical, Bramwell2001}. Recent research translated these concepts from magnetic spin systems to engineered soft-matter systems, such as acoustic channels~\cite{wang2017harnessing}, buckled elastic beams~\cite{kang2014complex}, and monolayers of colloidal spheres~\cite{han2008geometric, ShokefLubensky2009}. Here, frustration in magnetic spin systems is related to incompatibility of soft modes in mechanical systems~\cite{coulais2016, bertoldi2017flexible, meeussen2020topological, pisanty2021putting}. Mapping of spin systems to metamaterial architectures has provided insight into the mechanical consequences of frustration, such as stress control~\cite{meeussen2020response} and domain-wall topology~\cite{deng2020characterization, merrigan2021topologically}. In particular, exploring irreversibility and history dependence through these analogies has opened new routes for programmable elastic responses and mechanical memory storage~\cite{chen2021reprogrammable, bense2021, udani2021taming, guo2021non, shohat2021memory, jules2022delicate, lahini2017,shohat2023dissipation}, and has set the grounds for further advances in mechanical computing~\cite{Yasuda2021computing, treml2018origami}. However, multistability and complex memory formation are typically traits of disordered and amorphous systems~\cite{mungan2019networks, keim2020global, shohat2021memory}, which are harder to predict and control. On the other hand, periodic mechanical systems are generally not multistable, as their long range elastic interactions tend to resolve frustration with long-range ordered ground states~\cite{shokef-PNAS-2011, kang2014complex}. Here, the displacement of mechanical degrees of freedom can take intermediate values, leading to an ordered compromise and lifting the degeneracy associated with frustrated spin systems.

In this Article, we introduce {\it ordered} mechanical metamaterials that exhibit a large multiplicity of {\it disordered} metastable states. This leads to multistability of internal degrees of freedom and to mechanical memory, as configurations can depend on their preparation history. Crucially, our approach relies on spatial separation between the frustrated motifs in the metamaterial, which we achieve via a mapping to {\it vertex-frustrated}~\cite{morrison2013unhappy} artificial spin ice (ASI)~\cite{wang2006ASI}. As a result, long-range cooperative effects are suppressed, and disorder emerges from local rules. Furthermore, the graph of transitions between the metamaterial's states~\cite{mungan2019networks} contains irreversible pathways, and its structure leads to non-Abelian behavior. Namely, the system's state depends not only on the external manipulations applied on the metamaterial, but also on their precise sequence. We utilize this to demonstrate how the history of operations acted on the system, and their precise sequence, may be inferred from a static measurement of the system's final state. This takes an important step towards a systematic implementation of computation in mechanical metamaterials~\cite{ding2022sequential}.

The design of our mechanical metamaterial is inspired by the frustration-based designs of ASI~\cite{wang2006ASI, nisoli2013colloquium, skjaervo2020advances}, which lead to constrained disorder, and thus to a degenerate manifold of configurations often captured by interesting emergent descriptions~\cite{gilbert2016frustration, nisoli2017deliberate}. There are strong similarities between ASI and mechanical metamaterials realized by repeated arrangements of simple units endowed with a soft mode: In ASI, nano-islands are arranged along the edges of a lattice, and their magnetization is described by binary arrows pointing toward or from the vertices of the lattice, with certain vertex configurations locally minimizing the energy. Identically, in a mechanical metamaterial, displacements point into or out of the repeating units, and the softest deformation mode of each unit is related to a certain mutual arrangement of these displacements. 

\begin{figure}[t!]
\centering
\includegraphics[width=0.56\linewidth]{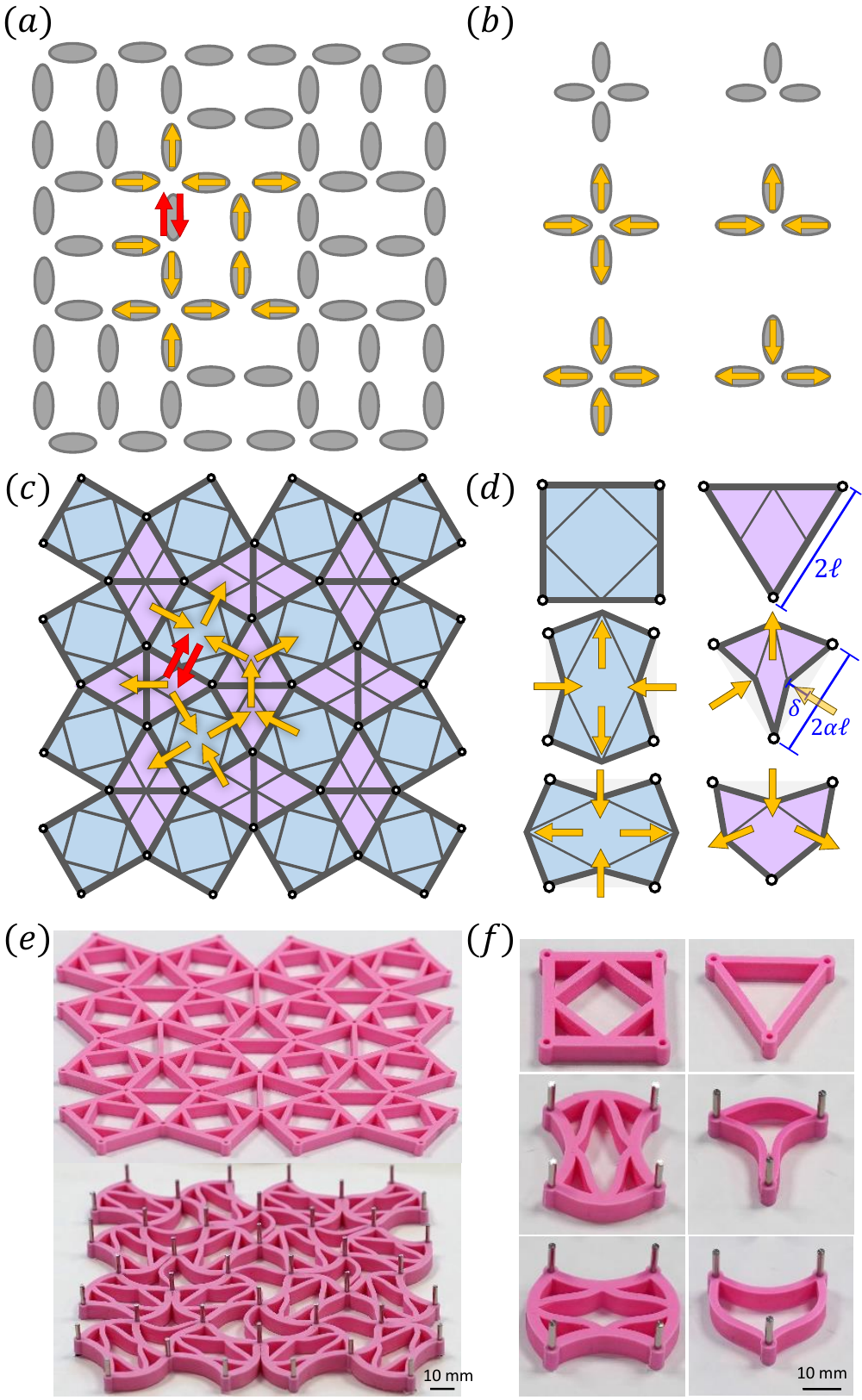}
\caption{\textbf{Design principle} - (a) The Shakti lattice artificial spin ice (ASI) exhibits vertex frustration: A conflict (red arrows) results when trying to place all vertices around a rectangular plaquette in their lowest-energy states (as individually shown in b). (c) The Chaco lattice mechanical metamaterial inherits this vertex frustration of the Shakti ASI: The five units surrounding any corner in the metamaterial may not simultaneously deform according to their softest mode (as individually shown in d). Thick lines indicate $k_1$ springs forming the edges of the square and triangular units. Thin lines indicate $k_2$ springs responsible for the interactions between neighboring edges within each unit. Spontaneous edge displacements are generated by pinning the corners of all units (white dots) to distances smaller than the relaxed edge length. (e) Experimental realization of the Chaco metamaterial, and (f) the softest deformations of its units. In the triangular units, the different stiffnesses at the corners connecting adjacent edges give rise to the softest modes obtained by the $k_2$ springs in the theoretical model.}
\label{fig:chaco_lattice}
\end{figure}

The similarity extends to frustration and incompatibility: In vertex-frustrated ASI lattices, not all vertices can simultaneously be in their lowest-energy state~\cite{morrison2013unhappy, stamps2014artificial}, as demonstrated for the Shakti-lattice ASI in Fig.~\ref{fig:chaco_lattice}(a). In such lattices, when going along a loop of vertices around any plaquette of the lattice, the spins may not be assigned in a way such that all vertices will be in their energetically-preferred configurations. Equivalently, in combinatorial mechanical metamaterials, one can arrange the units so that they cannot all simultaneously deform according to their soft mode~\cite{coulais2016, meeussen2020topological, pisanty2021putting}. Vertex frustration in magnetism thus corresponds to incompatibility in mechanical metamaterials. The plethora of frustrated ASI geometries and topologies~\cite{morrison2013unhappy, gilbert2014emergent, skjaervo2020advances, saglam2022entropy, rodriguez2023geometrical} inspires novel metamaterial designs. We show that in the mechanical system such ASI-inspired design can suppress long-range ordered ground states, and lead to emergent disorder and complex memory formation.


\section{The Chaco Mechanical Metamaterial}

In the Shakti ASI~\cite{morrison2013unhappy, chern2013degeneracy, gilbert2014emergent, gilbert2016frustration, lao2018classical}, vertices have coordination numbers (or number of impinging arrows) $z=2,3,4$. Vertices where $z=3$ or $z=4$ spins meet have two distinct energy-minimizing configurations [Fig.~\ref{fig:chaco_lattice}(b)], and $z=2$ vertices prefer to have their spins aligned. Because of the vertex-frustrated lattice geometry, it is impossible to set all vertices to these lowest-energy configurations [see rectangular loop (yellow) in Fig.~\ref{fig:chaco_lattice}(a)], without resulting in a conflict (red arrows). Excitations of the $z=3$ vertices cost less energy than excitations of the $z=4$ vertices. Consequently, the extensively-degenerate, disordered ground state of the Shakti ASI is achieved by any distribution of $z=3$ excitations such that each excitation resolves the frustration for two minimal loops of spins. The remaining $z=3$ vertices and all the $z=4$ vertices adopt minimum-energy configurations.

In Fig.~\ref{fig:chaco_lattice}(c) we introduce what we shall refer to as the Chaco-lattice mechanical metamaterial, whose incompatibility corresponds to the frustration of the Shakti ASI. The $z=3$ and $z=4$ vertices of the magnetic Shakti  [Fig.~\ref{fig:chaco_lattice}(b)] correspond to the triangular and square units, respectively, comprising the mechanical Chaco [Fig.~\ref{fig:chaco_lattice}(d)]. The arrows describing the magnetic moments in the lowest-energy states of the former correspond to the softest, or lowest-energy deformations of the latter. We model the mechanical Chaco as a network of linear springs. Each edge of the squares and triangles consists of two springs with stiffness $k_1$ and rest length $\ell$. These are connected by internal coupling springs of stiffness $k_2$. 

In ASI, nanoislands are naturally magnetized. In metamaterials, one can induce an equivalent spontaneous displacement of all edges by prestressing the system~\cite{merrigan2021topologically}. Namely, pinning an edge to a distance $2 \alpha \ell $, with $\alpha < 1$, imposes a compression, which causes it to buckle by an amount $\delta =\ell \sqrt{1-\alpha^2}$. This results in a contraction by a factor $\alpha^2$ of the two-dimensional lattice. In the prestressed Chaco lattice, the units may not all simultaneously adopt their zero-energy deformations, due to the inherent incompatibility. Nevertheless, in the weak coupling limit, $k_2 \ll k_1$, we expect the free nodes to behave in an approximately binary way~\cite{merrigan2021topologically}, adopting displacements $\vec{s}_i \approx \pm \delta \hat{z}_i$ from their rest position, where $\hat{z}_i$ is the local direction perpendicular to the edge. In this limit, the $k_1$ bonds are approximately relaxed and most of the strain is concentrated on the $k_2$ bonds. As we discuss below, the overall compression applied to the lattice breaks the symmetry between stretching and compressing, such that compressing a $k_2$ bond costs more energy than stretching it.

Experimentally, we create the elastic network shown in Fig.~\ref{fig:chaco_lattice}(e), top. Similarly to the spring model, a thin enough elastic beam of length $2\ell$ compressed by a factor $\alpha$ tends to buckle via its first deformation mode. We prestress the rubber network by constraining the corners of all square and triangular units to an array of metal pins connected to a rigid substrate, and spaced such that the network is uniformly compressed by factor $\alpha$ [Fig.~\ref{fig:chaco_lattice}(e), bottom]. The geometry of the units leads to mechanical soft modes [Fig.~\ref{fig:chaco_lattice}(f)], which reproduce those of the theoretical springs model [Fig.~\ref{fig:chaco_lattice}(d)].  See Appendix~\ref{app_experimental} for experimental details.

\section{Mechanical Equilibrium States}

\begin{figure}[h]
\centering
\includegraphics[width=0.52\linewidth]{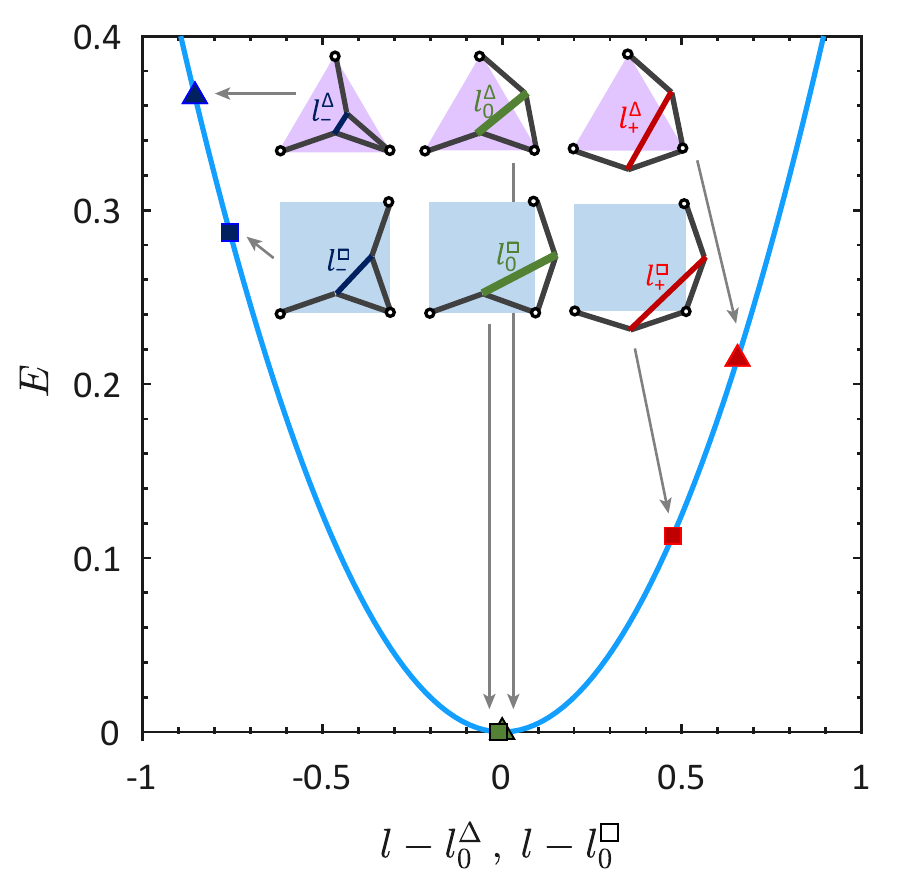}
\caption{\textbf{$Z_2$ symmetry breaking} - Energetics of the ideal nearest-neighbor interactions for the square and triangular mechanical units. Ideal displacements allow the $k_1$ bonds to be relaxed and place all energy on the internal $k_2$ bonds. Spin inversion symmetry is broken in the mechanical system: bond extension costs less energy than bond compression.}   
\label{fig:energy_parabola}
\end{figure}

The overall compression breaks the $Z_2$ spin-reversal symmetry between nearest-neighbor degrees of freedom. The mechanical degrees of freedom behave in a spin-like binary way for $k_2 \ll k_1$, whereas elastic, continuous deformations are expected otherwise. Consider the interaction energies when ideal displacements $\pm \delta$ are imposed on pairs of neighboring edges: One in and one out displacement allow the internal $k_2$ bonds to rotate and remain in their relaxed lengths $l_0^{\framebox(4,4){}} = \sqrt{2} \ell$ and $l_0^{\Delta} = \ell$. Two out displacements result in a stretched internal $k_2$ bond, with lengths $l^{\framebox(4,4){}}_{+} $ and $l^{\Delta}_{+}$, and energies $E^{\framebox(4,4){}}_+$ and $E^{\Delta}_+$. Two in displacements result in a compressed internal $k_2$ bond, with lengths $ l^{\framebox(4,4){}}_{-} $ and $l^{\Delta}_{-}$, and energies $E^{\framebox(4,4){}}_-$ and $E^{\Delta}_-$. Due to the overall compression, stretching the internal bonds costs less energy than compressing them further, as shown in Fig.~\ref{fig:energy_parabola}. An important consequence of this breaking of the $Z_2$ symmetry is that excited triangular and square mechanical units have distinct energy hierarchies than the $z=3$ and $z=4$ Shakti ASI vertices~\cite{morrison2013unhappy}. In the ideal binary limit, $k_2 \ll k_1$, the energy of a mechanical unit follows from the number of stretched and compressed bonds. There are six energy levels for the square units with energies: $0$, $2E^{\framebox(4,4){}}_{+}$, $E^{\framebox(4,4){}}_{+} + E^{\framebox(4,4){}}_{-}$, $2E^{\framebox(4,4){}}_{-}$, $4E^{\framebox(4,4){}}_{+}$, and $4E^{\framebox(4,4){}}_{-}$. Triangles have five possible energies, $0$, $E^{\Delta}_{+}$, $E^{\Delta}_{-}$, $2E^{\Delta}_{+}$, and $2E^{\Delta}_{-}$.

\begin{figure*}[h]
\centering
\includegraphics[width=0.8\linewidth]{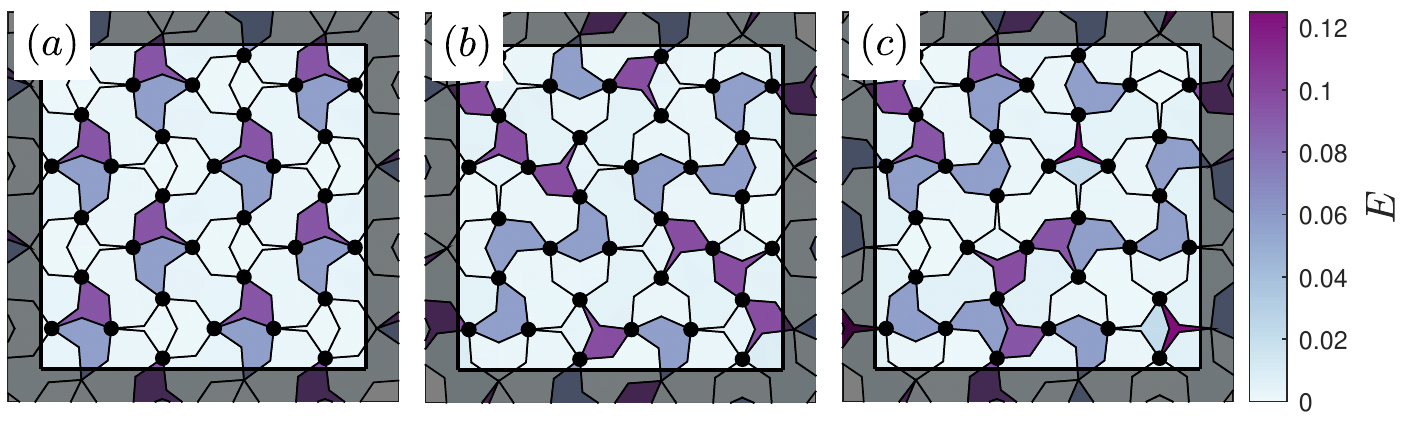}
\caption{\textbf{Multiple stable configurations} - Square and triangle unit energies for sample configurations with increasing energy in a $4\times4$ Chaco lattice with periodic boundary conditions, all for $k_2/k_1 = 0.033$ and $\alpha=0.9$: (a) Ordered ground state with energy localized on vertical pairs of triangles. (b) Complex stress distribution of a metastable state derived from a magnetic Shakti ground state, in which each stressed triangle is adjacent to a relaxed triangle. (c) Metastable configuration with all squares in relaxed random orientations and stress localized to more than half of the triangles.}
\label{fig:pbc_states}
\end{figure*}

Similarly to the magnetic Shakti, the most energy-efficient way to resolve the frustration in the mechanical Chaco is to localize stress on half of the triangles. This leaves the remaining triangles and all squares close to their zero-energy states. However, the continuous nature of the mechanical degrees of freedom allows minimizing the energy by long-range pairing of the excited triangles. Namely, within two adjacent excited triangles, the $k_2$ bonds do not have to take the ideal values $l^{\Delta}_{+}$ and $l^{\Delta}_{-}$ mentioned above. Instead, due to the nonlinear dependence of energy on displacement, there is an intermediate balance between their deformations which minimizes the energy. In a perfect realization, this leads to an ordered ground state with defect pairing [Fig.~\ref{fig:pbc_states}(a)], which has a four-fold degeneracy, since stresses may be placed on either vertical or horizontal pairs of triangles and since spin reversal leaves the energy unchanged. For sufficiently large $k_2/k_1$, the ground state is the only mechanically-stable configuration. However, for small $k_2/k_1$, we also find an extensive manifold of metastable states. Namely, any configuration with all squares in one of their individual zero-energy states becomes metastable. This holds for configurations with stresses localized to exactly half of the triangles, equivalent to any ground state of the Shakti [Fig.~\ref{fig:pbc_states}(b)]. It also holds for configurations where more than half of the triangles are excited [Fig.~\ref{fig:pbc_states}(c)]. Although these states have higher energy than the ground state, the excess stress in the triangles does not suffice to overcome the energy barrier for flipping the squares. Thus, since thermal fluctuations are negligible, if the system is in such a disordered metastable state, it will remain there and will not relax to the ordered ground state. See Appendix~\ref{app_computational} for computational details.

\section{Memory and Irreversible Dynamics}

\begin{figure*}[t]
\centering
\includegraphics[width=0.75\linewidth]{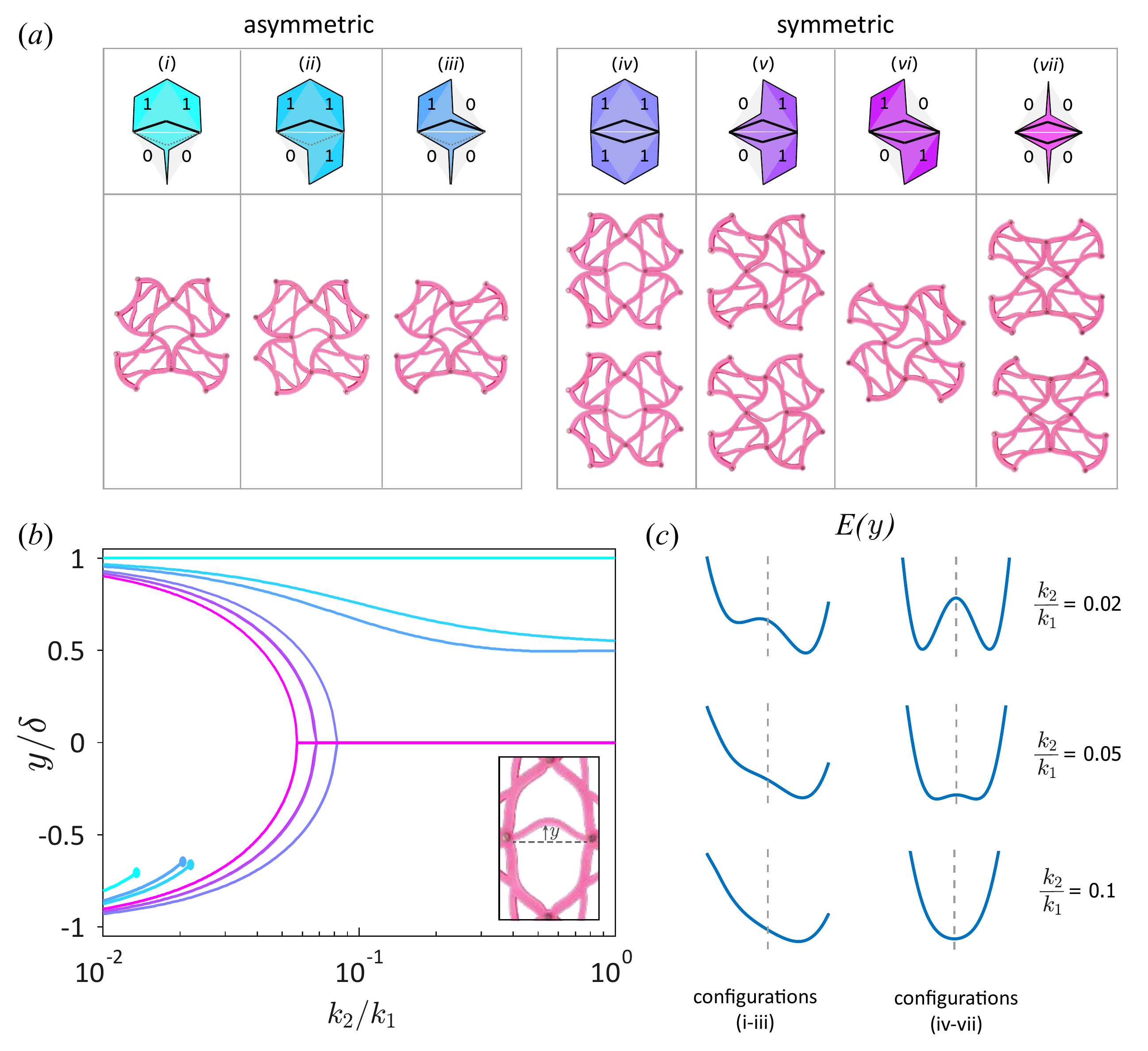}
\caption{\textbf{Stability of double triangles} - (a) The seven distinct configurations of a pair of back-to-back triangles and their surrounding four squares within the Chaco lattice, with 1 (0) indicating displacement out (in) of the triangle. We divide them by the symmetry of forces applied to the central beam; Configurations $i$-$iii$ are monostable. Configurations $iv$,$v$,$vii$ are bistable. Configuration $vi$ is bistable in the theoretical springs model, while experimentally its central beam exhibits a single stable state with a second buckling mode. (b) Stable vertical positions of the central point between the two triangles as a function of $k_2/k_1$ for $\alpha=0.9$ and assuming that the four external edges are fixed with ideal displacements $\vec{s}_i= \pm \delta \hat{z}_i$. Inset indicates the vertical displacement $y$. (c) Potential energy $E(y)$ as function of central beam displacement for configurations $ii$ (left) and $vi$ (right) representing the two classes. At high $k_2/k_1$ the symmetric configurations become monostable (bottom right). At low $k_2/k_1$ the asymmetric configurations develop metastability (top left). In between is our operational regime where all configurations follow the behavior shown in panel (a).}
\label{fig:double_triangle}
\end{figure*}

We distinguish between two types of degrees of freedom within the Chaco metamaterial -- the square units, and the central edges between pairs of back-to-back triangles. For small $k_2/k_1$, we describe both of them as approximately binary variables, stuck in one of their stable states. Namely, squares adopt one of their soft configurations, and each central edge buckles to one of two directions. The latter degrees of freedom constitute \textit{hysterons}~\cite{keim2019memory, bense2021, van2021profusion}, bistable elements whose state is history dependent. Hence, they are a natural representation for memory in the metamaterial. As described above, both states of the squares are always stable. In contrast, the stability of the central edge between two triangles depends on the configuration of the four squares surrounding it. We label the $2^4=16$ possible states of these squares by $2\times2$ matrices with binary entries, where $0$ indicates a displacement into the double triangle and $1$ out of it. By symmetry, these $16$ states reduce to the seven distinct configurations shown in Fig.~\ref{fig:double_triangle}(a).

We now fix all squares to their zero-energy states, and compute the minimal-energy states of the central edge as function of $k_2/k_1$, for each configuration. The horizontal displacement of the central point is negligible, thus we focus on its vertical displacement $y$. We find that it may exhibit monostable, metastable or bistable behavior for the different configurations and for different values of $k_2/k_1$, as shown in Fig.~\ref{fig:double_triangle}(b,c): Due to their vertical asymmetry, configurations $i-iii$ are typically monostable, with $y>0$: configuration $i$ allows both triangles to adopt zero-energy states, such that $y=\delta$ for any $k_2/k_1$; configurations $ii,iii$ show a reduced amplitude $0 < y < \delta$, which decreases as $k_2/k_1$ increases. For very small values of $k_2/k_1$, metastable solutions with $y<0$ appear for configurations $i-iii$. Configurations $iv-vii$ are vertically symmetric. As a result, the central edge is bistable for small $k_2/k_1$, while for larger $k_2/k_1$, the central point adopts the value $y=0$. From here on, we focus on $0.022 \lesssim k_2/k_1 \lesssim 0.057$, where configurations $i-iii$ are strictly monostable and configurations $iv-vii$ are bistable, as shown in Fig.~\ref{fig:double_triangle}(c), middle panel.

Our experimental design is similarly aimed to exhibit such behavior for all seven configurations.  However, as opposed to the theoretical spring model, in the experimental system, the squares apply torque to the edges of the central beam. In configuration $vi$ the torques are not symmetric under up-down reflection, and consequentially they excite the second bending mode of the central beam, as shown in Fig. \ref{fig:double_triangle}a. Hence, this configuration is bistable only in the theoretical springs model, while it is monostable in the experimental network of elastic beams.

\begin{figure*}[b]
\centering
\includegraphics[width=0.9\linewidth]{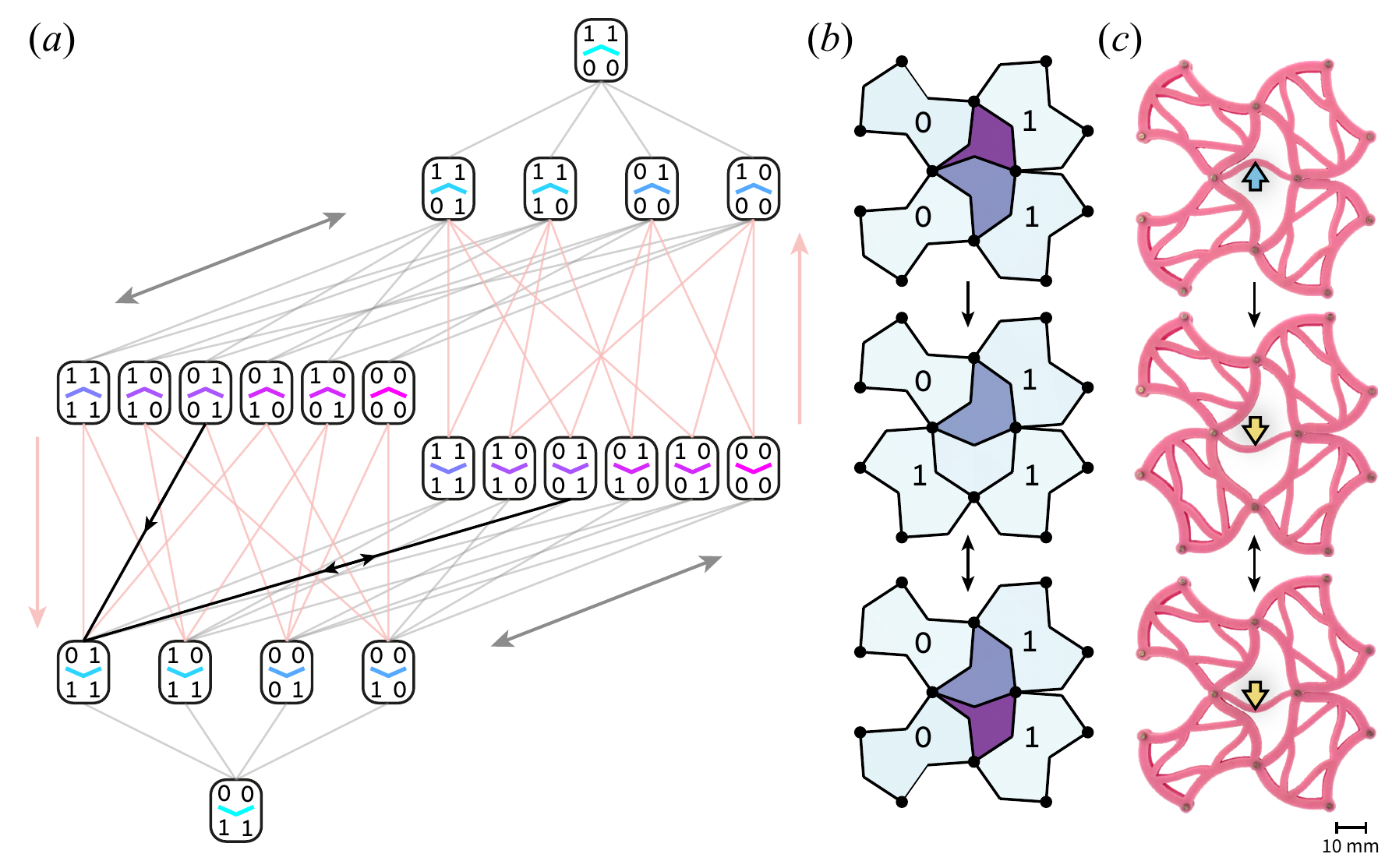}
\caption{\textbf{Local memory and irreversibility} - (a) Transition graph between the possible states. Bistable configurations are shown twice to include the possible deflections of the central edge. Reversible transitions are plotted in grey. Irreversible transitions, which flip the central edge, are plotted in peach. (b,c) Irreversible control sequence for the internal bistable edge in simulations (b) and experiment (c). The starting configuration (top) is bistable with the internal edge up (cyan arrow). Flipping the lower-left square (middle) creates a monostable configuration, so the internal edge flips down (yellow arrow). The lower-left square is returned to its original bistable state (bottom), but the internal edge remains pointing down. The pathway corresponding to panels (b,c) is highlighted in black on the transition graph (a).}
\label{fig:transition_graph}
\end{figure*} 

Since the configuration of the squares governs the stability of the central edge, we can precisely control the states of the triangles by manipulating the squares around them. Interestingly, the final state of the central edge depends not only on the configuration of squares, but on the exact order in which they were flipped, and consequently transition pathways are sequence dependent; The transitions between states are best understood using the directed graph~\cite{mungan2019networks, bense2021, van2021profusion} shown in Fig.~\ref{fig:transition_graph}(a), where we separately mark the two possible states of the central edge for the six bistable configurations, resulting in a total of 22 states. All transitions between states with the same direction of the central edge are reversible (gray). An irreversible transition (peach) occurs whenever a bistable configuration is connected to a monostable configuration with the opposite direction of the central edge. 

We propose a protocol, shown in simulations and experiments in Fig.~\ref{fig:transition_graph}(b,c), respectively, that demonstrates the aforementioned control: We start from the bistable configuration $\left( \begin{smallmatrix} 0 & 1 \\ 0 & 1 \end{smallmatrix} \right)$, with the central edge pointing upwards (top panels). Next, we flip the lower-left square, thus switching to the monostable state $\left( \begin{smallmatrix} 0 & 1 \\ 1 & 1 \end{smallmatrix} \right)$. This indirectly forces the central edge to flip downwards (middle panels). Finally, we flip the lower-left square back to its initial state, which brings us back to the original bistable configuration $\left( \begin{smallmatrix} 0 & 1 \\ 0 & 1 \end{smallmatrix} \right)$. Interestingly, the central edge does not revert back, but remains down (bottom panels). After this irreversible path, if we continue flipping the lower-left square, the system reversibly switches back and forth between the latter two states.

\section{Non-Abelian Response}

\begin{figure*}[h]
\centering
\includegraphics[width=0.45\linewidth]{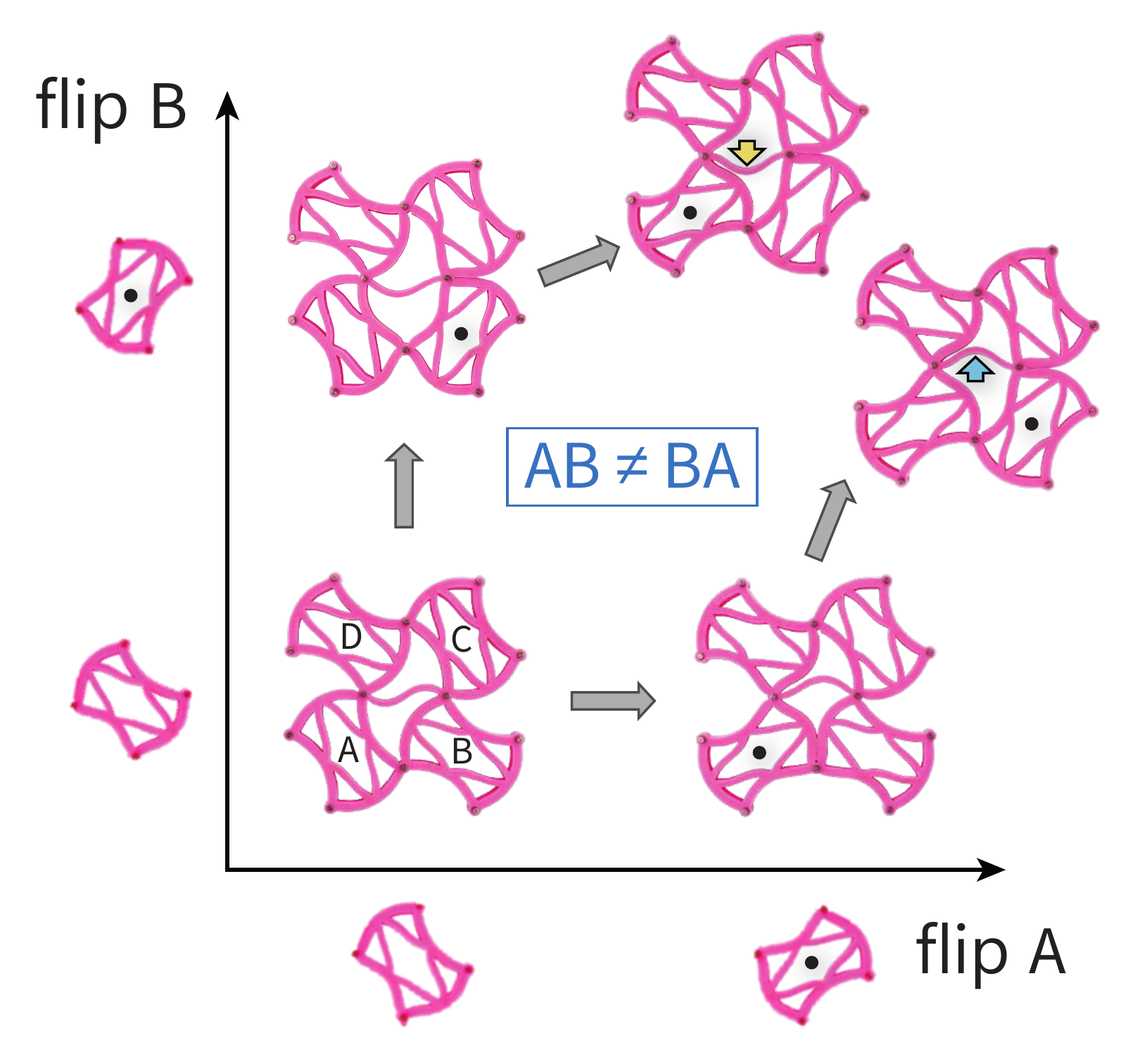}
\caption{\textbf{Non-Abelian response} - starting from state $\left(\begin{smallmatrix} 0 & 1 \\ 1 & 0 \end{smallmatrix} \right)$, consider flipping the bottom squares marked $A$ and $B$ as two operations acting on the system. The response of the Chaco metamaterial to these operations depends on the order in which they are applied, as indicated by the displacement of the central edge. Black dots mark the square that has just flipped.}
\label{fig:non_Abelian_small}
\end{figure*}

The subsystem of $2 \times 2$ squares analyzed above suffices to give rise to rich dynamics and pathways within the Chaco metamaterial. The generic motif underlying this richness, we find, is path dependency. The state of the Chaco metamaterial is sensitive to the ordering of external operations, rendering it non-Abelian. Recently, such non-Abelian responses were used to realize both finite state machines~\cite{singh2023emergent} and Set-Reset latches~\cite{guo2021non}, which are elementary components of computation. To demonstrate this here, we start from the state $\left( \begin{smallmatrix} 0 & 1 \\ 1 & 0 \end{smallmatrix} \right)$, and consider the response to flips of the surrounding squares $A,B,C,D$, as shown in Fig.~\ref{fig:non_Abelian_small}. Upon flipping squares $A$ and $B$, the final state depends on the flipping sequence. In other words, the operators $A$ and $B$ of flipping these two squares do not commute, $AB\neq BA$.

\section{Sequence Recognition}

Next, we consider an extended Chaco lattice with $4 \times 4$ squares. Here, the dynamics of each double triangle is governed by the transition graph shown in Fig.~\ref{fig:transition_graph}(a). We can thus understand the dynamics of a large system, based on the local rules described above. Using four different inputs, we now show that the generic non-Abelian response presented in Fig~\ref{fig:non_Abelian_small} can be harnessed to implement sequence recognition. Namely, after a set of square flips, the state of the double triangles may allow to infer the order of these flips and to detect a chosen sequence.

\begin{figure*}[b]
\centering
\includegraphics[width=1\linewidth]{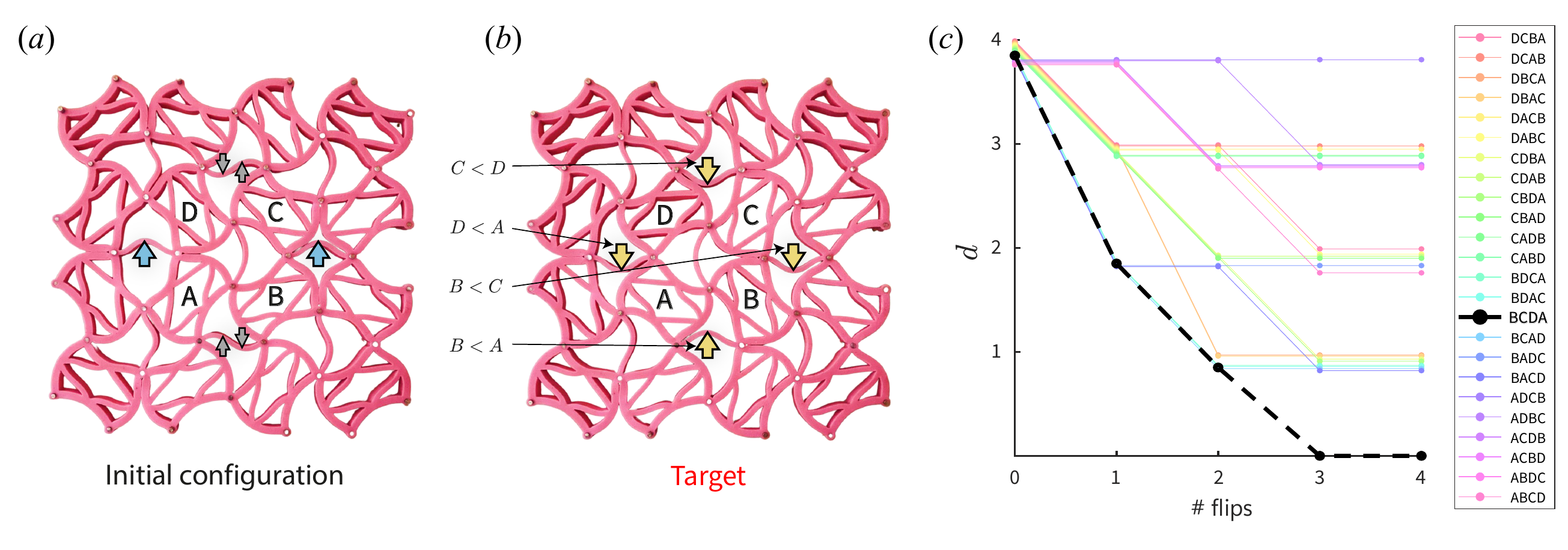}
\caption{\textbf{Sequence Recognition} - (a) Initial configuration for the sequence recognition protocol, where the squares marked $A-D$ are flipped; (b) Target configuration after all squares have been flipped. The state of the double triangles, denoted $\boldsymbol{\sigma}^{0}$ translates to a set of inequalities for the flipping order; (c) distance from the target $d=\sum_{i}\left(\sigma_{i}-\sigma^{t}_{i}\right)^{2}$ along any flip sequence. The sequence $BCDA$ which satisfies all inequalities is the only one reaching the target, and the distance metric allows sequence recognition.}
\label{fig:seq_det}
\end{figure*} 

Starting from the initial configuration shown in Fig.~\ref{fig:seq_det}a, we flip the four middle squares marked $A,B,C,D$. After flipping all four squares, the double triangles shared by each two squares encode their flipping sequence. In other words, the final state of the four relevant double triangles (marked by arrows), encodes the partial ordering of the flips of each pair of adjacent squares. Since this ordering is partial, it is insensitive to switching the order of non-neighboring flips (for example $A$ and $C$, or $B$ and $D$). Therefore, we can discern between $2^4=16$ out of the total $4!=24$ possible sequences.

We now consider a target configuration, for instance the one shown in Fig.~\ref{fig:seq_det}b, and denote it~$\boldsymbol{\sigma}^{t}$, the state vector of all double triangles. We ask which flip sequence results in this configuration. The partial ordering between neighboring flips can be translated to a series of inequalities. Solving these inequalities allows us to analytically obtain the desired sequence, in this case $BCDA$. We can now formulate a metric $d=\sum_{i}\left(\sigma_{i}-\sigma^{t}_{i}\right)^{2}$ which captures the distance of a given state from the target configuration. Following $d$ along any flip sequence reveals that indeed, only the theoretically predicted sequence $BCDA$ leads to the target, as shown in Fig.~\ref{fig:seq_det}c. All other sequences end in final configurations, which differ from the target configuration. This metric can implement sequence recognition. By following $d$ one can infer whether the sequence matched the target ($d=0$) or was incorrect ($d>0$) as shown in Fig.~\ref{fig:seq_evolution} and in the supplemental movie~\cite{supplemental}, for the correct sequence and for an example of an incorrect sequence.

\begin{figure*}[h]
\centering
\includegraphics[width=1\linewidth]{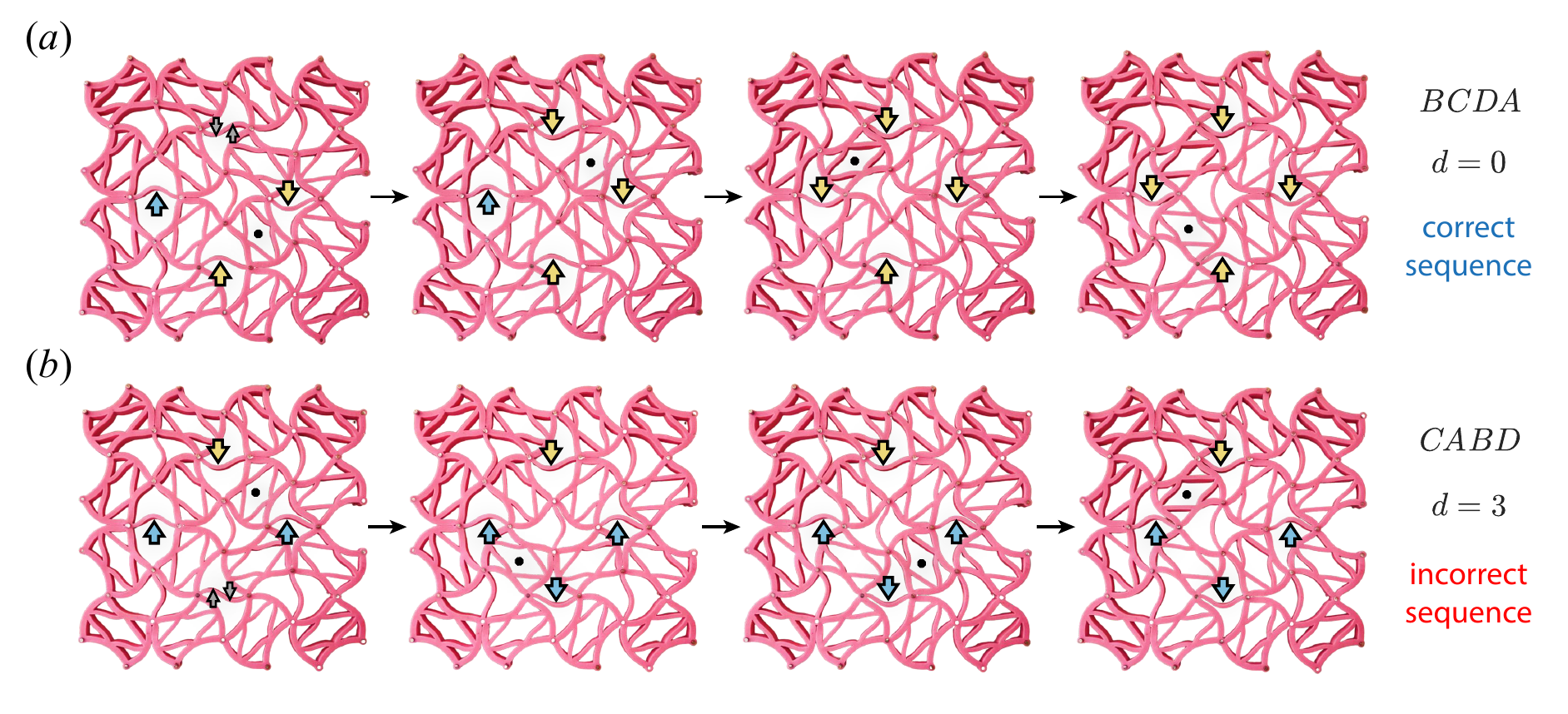}
\caption{\textbf{State Evolution During Flip Sequences} - Starting from the initial condition of Fig.~\ref{fig:seq_det}a, we track the state of the system along two sequences. Black dots mark the square that was flipped in each stage. The correct sequence $BCDA$ (a) leads to the target state, or $d=0$; in contrast, an incorrect sequence $CABD$ results in the wrong state, measured by the metric $d=3$.}
\label{fig:seq_evolution}
\end{figure*} 

\section{Discussion}

The mechanism underlying the non-Abelian response sheds light on the generic emergence of sequence-dependent responses in frustrated mechanical systems~\cite{Fruchart_2020_dualities, udani2021taming, liu2021frustrating, guo2021non, hexner2021training}. The order in which bistable degrees of freedom are manipulated follows a pathway of states, which favors one of the possible final states. In essence, the sequence directs the otherwise spontaneous symmetry breaking of the system at its final state. This framework may be used to cleverly manipulate mechanical systems between multiple functionalities~\cite{coulais2018multi, yuan2021recent, bossart2021oligomodal}, as different states of the system may exhibit different mechanical responses. This, we envision, may allow adaptable, history dependent mechanical responses via \textit{in materia} information processing \cite{treml2018origami,Yasuda2021computing}.

We have focused theoretically and experimentally on the weak interaction limit (measured by the coupling parameter $k_{2}/k_{1}$). Strong interactions between hysterons lead to long-range order and a distinct ordered ground state~\cite{kang2014complex}, see Appendix~\ref{app_collapse}. However, intermediate interaction strengths may give rise to even richer dynamics~\cite{shohat2021memory, bense2021}. For example, they can lead to long transients, or complex cycles with longer periods~\cite{lindeman2021multiple, keim2021multiperiodic, Szulc_JCP_2022}, which can realize an array of computational tasks such as counting~\cite{kwakernaak2023counting}.

Following the vertex-frustration analogy back to the magnetic realm, non-Abelian protocols may increase the memory capacity in storage devices based on frustrated geometries. Due to the sequence sensitivity, manipulating $n$ binary degrees of freedom may encode more than $2^n$ unique states, as the system's history may influence additional features that are beyond the states of the manipulated degrees of freedom. Thus the manifold of reachable states may be enriched by frustrated interactions. 

Finally, the Chaco mechanical metamaterial inherits from the Shakti ASI a topological structure: the possible allocation of its incompatibilities can be mapped both in the free fermion point of a six vertex model~\cite{chern2013degeneracy} and thus into a dimer cover model~\cite{lao2018classical}. These height models can be thought of as instances of so-called classical topological order~\cite{lamberty2013classical} for which boundary conditions strongly constrain the manifold in the bulk, and in the case of Chaco metamaterial, that would imply an encoding via the boundaries. Moreover, using the same dimer mapping and the design strategy by triangle rotation pioneered in ref~\cite{meeussen2020topological} for the kagome lattice, an extensive number of Chaco-based metamaterials can be produced.  

\newpage 

\begin{acknowledgements}

We thank Ben Pisanty, Eial Teomy, Erdal O\u{g}uz, Izhar Neder, Muhittin Mungan, Priyanka, Roni Ilan, Yael Roichman, and Ofer Shochet for fruitful discussions. This research was supported in part by the Israel Science Foundation Grants No. 1899/20 (Y.S.) and 2117/22 (Y.L.). Y.S. and C.M. thank the Center for Nonlinear Studies at Los Alamos National Laboratory for its hospitality. The work of C.N. was carried out under the auspices of the U.S. DoE through the Los Alamos National Laboratory, operated by Triad National Security, LLC (Contract No.229892333218NCA000001) and founded by a grant from the DOE-LDRD office.

\end{acknowledgements}

\clearpage

\appendix

\section{Experimental Methods}
\label{app_experimental}

\begin{figure}[h]
\centering
\includegraphics[width=0.4\linewidth]{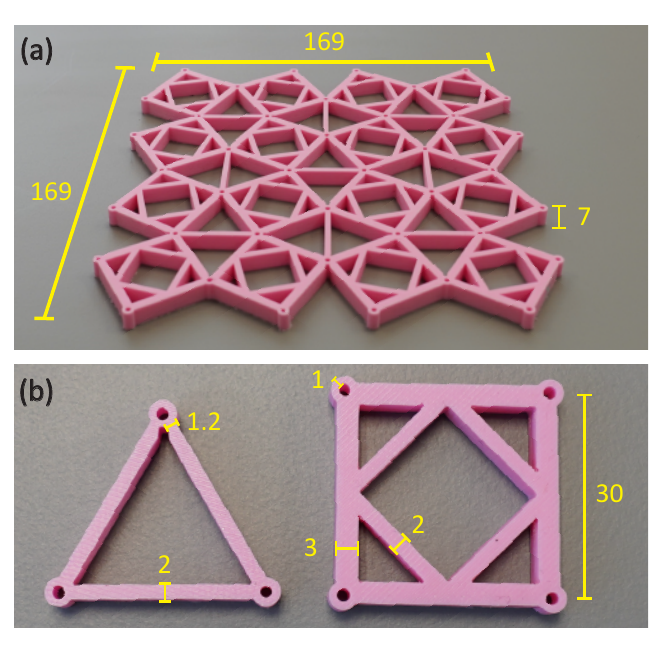}
\caption{Geometric dimensions (in~$mm$) of the experimental Chaco lattice.}
\label{fig:exp_design}
\end{figure}

To realize the frustrated Chaco lattice, we design a network of elastic beams, tracing the geometry of the Hookean springs in our theoretical model. Using a 3D printer (Prusa MK3 i3) we manufacture molds for the network, which are then used to cast samples made from silicone rubber (Mold Max™) with a Shore hardness of 30 A. The geometric dimensions of the rubber samples are shown in Fig.~\ref{fig:exp_design}. 

The basic beam length, constituting half the edge of each square or triangular unit in the Chaco lattice is equal to $\ell = 15mm$. To induce prestress in the samples, we constrain the elastic network to dimensions smaller than its rest configuration, using a set of pins, $2\pm0.01 mm$ in diameter. Each pin acts as a rotation axis fixed in space. The spacing between neighboring pins is set to induce a uniform compression factor of $\alpha=0.92\pm0.01$. To limit the deformation of the central beam between two triangles to its first buckling mode, we remove from the triangles the elastic beams which trace the $k_2$ springs. Instead, the region between squares meeting at a vertex of the lattice serve as a coupling mechanism between the deformation of adjacent beams in each triangle. The $k_2$ springs are not removed from the squares, however, to ensure the squares do not exhibit additional metastable states. The control sequence protocols are performed manually. The beams corresponding to the $k_2$ springs allow easy manipulation of the squares between their two stable states. We force the squares to flip by pushing one of their outer beams. See supplemental movie~\cite{supplemental}. Experiments are documented using a digital camera (Sony alpha3). In Figs.~\ref{fig:double_triangle}, \ref{fig:transition_graph} and~\ref{fig:non_Abelian_small}, the background is digitally removed for clarity.

\section{Computational Methods}
\label{app_computational}

We simulate the harmonic springs theoretical model for the Chaco metamaterial with overdamped dynamics, or the method of steepest energy descent. The equation of motion for a given point with position $\vec{r}_i$ is given by \begin{equation} \frac{d\vec{r}_i}{dt} = \frac{1}{\gamma}\sum_{\langle ij \rangle} -k_{ij}(e_{ij}-l_{ij})\hat{r}_{ij}, \end{equation} where the spring connecting point $i$ to point $j$ has spring constant $k_{ij}$, relaxed length $l_{ij}$, and current extension $e_{ij}=|\vec{r}_i - \vec{r}_j|$, and $\gamma$ is a linear drag coefficient, which sets the relaxation time scale in the system, which, we set to $\gamma/k_1 = 1$. However all the results we present are for the final state after the system has fully relaxed, thus this time scale is not relevant for the results we present. Energy is minimized by integrating the equations of motion with variable time step $\Delta t = 0.01-0.1$. 

For Figs.~\ref{fig:pbc_states} and \ref{fig:double_triangle}, we define the energy contained within each of the mechanical units, accounting for contributions from both $k_1$ and $k_2$ bonds, so that summing up all unit energies gives the total lattice energy. Each $k_1$ bond in the Chaco lattice is shared between two mechanical units, either between a square and a triangle or between two triangles. All the $k_2$ bonds are internal to a single unit. Thus we define the energy of each unit as the sum of harmonic spring energies of all its internal $k_2$ bonds plus one half the harmonic spring energy for each $k_1$ bond along the boundary of the unit. 

\section{Multistability and collapse to the ground state}
\label{app_collapse}

\begin{figure*}[b]
\centering
\includegraphics[width=0.8\linewidth]{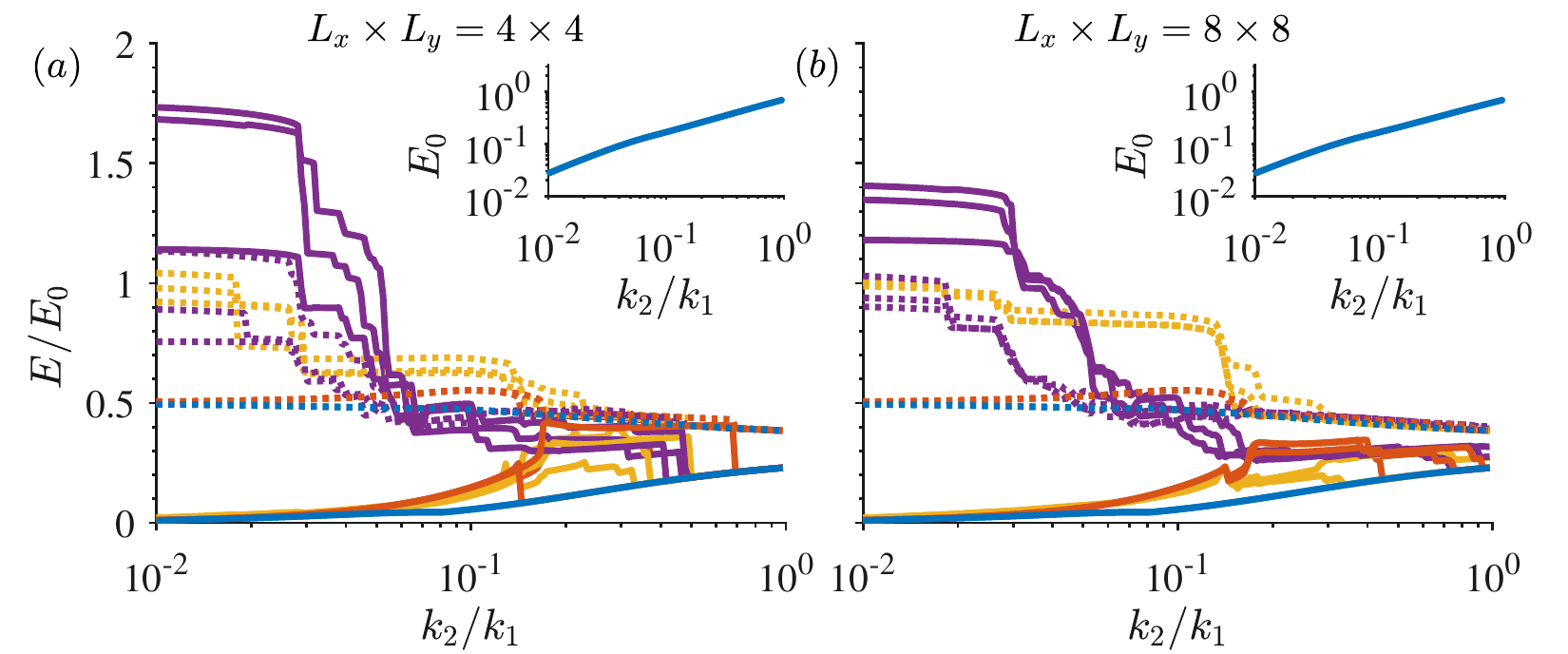}
\caption{Relative energy of the squares, $E^{\framebox(4,4){}} / E_0$ (solid) and of the triangles, $E^{\Delta} / E_0$ (dotted), both normalized by the average unit energy in the ground state $E_0 = E^{\framebox(4,4){}} + 2 E^{\Delta}$ (shown in the insets) as $k_2/k_1$ is slowly increased for systems of $4\times4$ (a) and $8\times8$ (b) squares with periodic boundary conditions. The line colors correspond to four classes of initial state: mechanical ground-state (blue), Shakti ASI ground-state with unpaired defects (red), randomly-oriented relaxed squares (yellow), and random displacements (purple). Three distinct initial conditions are shown for each class to illustrate the stochastic variations between runs. In (b), not all runs collapse to the ground state curves as $k_2/k_1 $ increases because of competing ground-state domains in the larger system.}
\label{fig:energy_curves}
\end{figure*}

Figure~\ref{fig:energy_curves} compares the relative energy contributions from the square and triangular units as $k_2/k_1$ is gradually increased starting from four classes of initial configuration: the mechanical ground state [Fig.~\ref{fig:pbc_states}(a)], a configuration derived from a Shakti-lattice ASI ground state without paired defects [Fig.~\ref{fig:pbc_states}(b)], a state with randomly chosen relaxed orientations for all the squares [Fig.~\ref{fig:pbc_states}(c)], and finally a state with random displacements $\vec{s}_i(0)$ of all edges. Once the starting metastable state becomes unstable, some edges within the lattice flip, creating sharp drops in the energy.

In the mechanical ground state [Fig.~\ref{fig:pbc_states}(a)], energy is localized to the triangles, with full localization $E^{\Delta}/E_0 \rightarrow 1/2 $ as $ k_2/k_1 \rightarrow0$. This behavior is also seen for the Shakti-lattice ASI ground state [Fig.~\ref{fig:pbc_states}(b)], which implies that the energy is also localized to the triangles, and further implies that the gap between the Shakti ASI ground state and the true mechanical ground state closes as $k_2/k_1 \rightarrow 0$. For larger $k_2/k_1$ the triangles and the squares in the Shakti ASI ground state have higher energy compared to the mechanical ground state. The energy of the squares in the Shakti ground state increases smoothly until $k_2/k_1 \approx 0.15$, where sharp jumps start to appear, indicating that some squares flip so that excited triangles can become paired.

For relaxed squares in random orientations [Fig.~\ref{fig:pbc_states}(c)], the energy is also localized to the triangles, but the presence of higher-order excitations on the triangles causes the energy stored to be nearly twice that of the triangle energies in the mechanical ground state, giving $E^{\Delta} / E_0 \rightarrow 1$ as $k_2/k_1 \rightarrow 0$. Finally, the state with random edge displacements has large energy stored in both the squares and the triangles. As $k_2/k_1$ increases, edges become unstable and flip, taking the configuration towards the ordered ground state. The three metastable states shown all reach the ground state for $k_2/k_1 > 0.5$. In larger systems, the true ground state is not always fully reached because several domains with different ground-state orientations may form.

\begin{figure*}[h]
\centering
\includegraphics[width=1\linewidth]{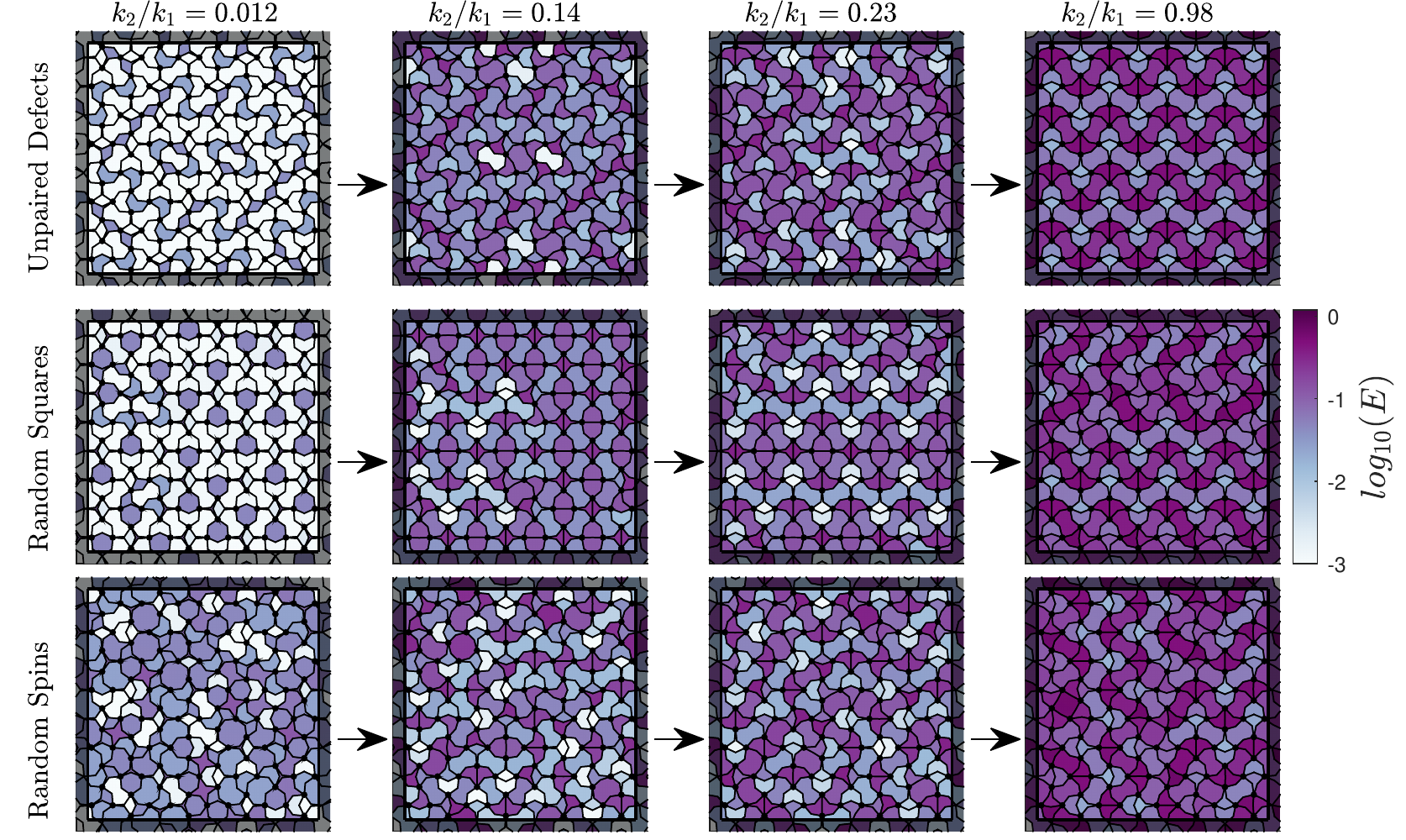}
\caption{Real space evolution of the metamaterial configurations for a system of $8\times8$ squares, for the three classes of initial condition, unpaired defects corresponding to a magnetic Shakti ground state, random initial choice of squares in their minimum energy orientations, and random choice of the edge displacements. As $k_2/k_1$ increases, edges inside the system start to flip to bring the configuration closer to the mechanical ground state. However, for the $k_2/k_1$ values shown here, there are still several competing ground-state domains, separated by domain walls.}
\label{fig:realspace8x8}
\end{figure*}

Figure~\ref{fig:energy_curves} plots the relative energy curves for two system sizes and also shows three separate initial conditions for the different classes considered. Other than stochastic variations in the positions of the energy jumps, due to different initial conditions, the intensive relative energy curves show the same behaviors for each class. The important major difference that appears in the larger system is some of the energy curves do not collapse to merge with the mechanical ground state curves as $k_2/k_1$ increases. Inspection of the real space evolution of the configurations for the larger system in Fig.~\ref{fig:realspace8x8} shows that the failure to collapse completely to the ground-state curves is a result of the presence of multiple competing ground-state domains.

\clearpage

\bibliography{chaco}

\begin{thebibliography}{57}%
\makeatletter
\providecommand \@ifxundefined [1]{%
 \@ifx{#1\undefined}
}%
\providecommand \@ifnum [1]{%
 \ifnum #1\expandafter \@firstoftwo
 \else \expandafter \@secondoftwo
 \fi
}%
\providecommand \@ifx [1]{%
 \ifx #1\expandafter \@firstoftwo
 \else \expandafter \@secondoftwo
 \fi
}%
\providecommand \natexlab [1]{#1}%
\providecommand \enquote  [1]{``#1''}%
\providecommand \bibnamefont  [1]{#1}%
\providecommand \bibfnamefont [1]{#1}%
\providecommand \citenamefont [1]{#1}%
\providecommand \href@noop [0]{\@secondoftwo}%
\providecommand \href [0]{\begingroup \@sanitize@url \@href}%
\providecommand \@href[1]{\@@startlink{#1}\@@href}%
\providecommand \@@href[1]{\endgroup#1\@@endlink}%
\providecommand \@sanitize@url [0]{\catcode `\\12\catcode `\$12\catcode
  `\&12\catcode `\#12\catcode `\^12\catcode `\_12\catcode `\%12\relax}%
\providecommand \@@startlink[1]{}%
\providecommand \@@endlink[0]{}%
\providecommand \url  [0]{\begingroup\@sanitize@url \@url }%
\providecommand \@url [1]{\endgroup\@href {#1}{\urlprefix }}%
\providecommand \urlprefix  [0]{URL }%
\providecommand \Eprint [0]{\href }%
\providecommand \doibase [0]{https://doi.org/}%
\providecommand \selectlanguage [0]{\@gobble}%
\providecommand \bibinfo  [0]{\@secondoftwo}%
\providecommand \bibfield  [0]{\@secondoftwo}%
\providecommand \translation [1]{[#1]}%
\providecommand \BibitemOpen [0]{}%
\providecommand \bibitemStop [0]{}%
\providecommand \bibitemNoStop [0]{.\EOS\space}%
\providecommand \EOS [0]{\spacefactor3000\relax}%
\providecommand \BibitemShut  [1]{\csname bibitem#1\endcsname}%
\let\auto@bib@innerbib\@empty
\bibitem [{\citenamefont {Ramirez}(1994)}]{ramirez1994strongly}%
  \BibitemOpen
  \bibfield  {author} {\bibinfo {author} {\bibfnamefont {A.~P.}\ \bibnamefont
  {Ramirez}},\ }\bibfield  {title} {\bibinfo {title} {Strongly geometrically
  frustrated magnets},\ }\href@noop {} {\bibfield  {journal} {\bibinfo
  {journal} {Annual Review of Materials Science}\ }\textbf {\bibinfo {volume}
  {24}},\ \bibinfo {pages} {453} (\bibinfo {year} {1994})}\BibitemShut
  {NoStop}%
\bibitem [{\citenamefont {Moessner}\ and\ \citenamefont
  {Ramirez}(2006)}]{moessner2006geometrical}%
  \BibitemOpen
  \bibfield  {author} {\bibinfo {author} {\bibfnamefont {R.}~\bibnamefont
  {Moessner}}\ and\ \bibinfo {author} {\bibfnamefont {A.~P.}\ \bibnamefont
  {Ramirez}},\ }\bibfield  {title} {\bibinfo {title} {Geometrical
  frustration},\ }\href@noop {} {\bibfield  {journal} {\bibinfo  {journal}
  {Physics Today}\ }\textbf {\bibinfo {volume} {59}},\ \bibinfo {pages} {24}
  (\bibinfo {year} {2006})}\BibitemShut {NoStop}%
\bibitem [{\citenamefont {Wannier}(1950)}]{wannier1950antiferromagnetism}%
  \BibitemOpen
  \bibfield  {author} {\bibinfo {author} {\bibfnamefont {G.~H.}\ \bibnamefont
  {Wannier}},\ }\bibfield  {title} {\bibinfo {title} {Antiferromagnetism. the
  triangular {I}sing net},\ }\href@noop {} {\bibfield  {journal} {\bibinfo
  {journal} {Physical Review}\ }\textbf {\bibinfo {volume} {79}},\ \bibinfo
  {pages} {357} (\bibinfo {year} {1950})}\BibitemShut {NoStop}%
\bibitem [{\citenamefont {Harris}\ \emph {et~al.}(1997)\citenamefont {Harris},
  \citenamefont {Bramwell}, \citenamefont {McMorrow}, \citenamefont {Zeiske},\
  and\ \citenamefont {Godfrey}}]{harris1997geometrical}%
  \BibitemOpen
  \bibfield  {author} {\bibinfo {author} {\bibfnamefont {M.~J.}\ \bibnamefont
  {Harris}}, \bibinfo {author} {\bibfnamefont {S.~T.}\ \bibnamefont
  {Bramwell}}, \bibinfo {author} {\bibfnamefont {D.~F.}\ \bibnamefont
  {McMorrow}}, \bibinfo {author} {\bibfnamefont {T.}~\bibnamefont {Zeiske}},\
  and\ \bibinfo {author} {\bibfnamefont {K.~W.}\ \bibnamefont {Godfrey}},\
  }\bibfield  {title} {\bibinfo {title} {Geometrical frustration in the
  ferromagnetic pyrochlore \ce{Ho2Ti2O7}},\ }\href@noop {} {\bibfield
  {journal} {\bibinfo  {journal} {Physical Review Letters}\ }\textbf {\bibinfo
  {volume} {79}},\ \bibinfo {pages} {2554} (\bibinfo {year}
  {1997})}\BibitemShut {NoStop}%
\bibitem [{\citenamefont {Bramwell}\ and\ \citenamefont
  {Gingras}(2001)}]{Bramwell2001}%
  \BibitemOpen
  \bibfield  {author} {\bibinfo {author} {\bibfnamefont {S.~T.}\ \bibnamefont
  {Bramwell}}\ and\ \bibinfo {author} {\bibfnamefont {M.~J.~P.}\ \bibnamefont
  {Gingras}},\ }\bibfield  {title} {\bibinfo {title} {Spin ice state in
  frustrated magnetic pyrochlore materials},\ }\href@noop {} {\bibfield
  {journal} {\bibinfo  {journal} {Science}\ }\textbf {\bibinfo {volume}
  {294}},\ \bibinfo {pages} {1495} (\bibinfo {year} {2001})}\BibitemShut
  {NoStop}%
\bibitem [{\citenamefont {Wang}\ \emph {et~al.}(2017)\citenamefont {Wang},
  \citenamefont {Zheng}, \citenamefont {Fernandes}, \citenamefont {Sun},
  \citenamefont {Xu}, \citenamefont {Sun}, \citenamefont {Kang}, \citenamefont
  {Tournat},\ and\ \citenamefont {Bertoldi}}]{wang2017harnessing}%
  \BibitemOpen
  \bibfield  {author} {\bibinfo {author} {\bibfnamefont {P.}~\bibnamefont
  {Wang}}, \bibinfo {author} {\bibfnamefont {Y.}~\bibnamefont {Zheng}},
  \bibinfo {author} {\bibfnamefont {M.~C.}\ \bibnamefont {Fernandes}}, \bibinfo
  {author} {\bibfnamefont {Y.}~\bibnamefont {Sun}}, \bibinfo {author}
  {\bibfnamefont {K.}~\bibnamefont {Xu}}, \bibinfo {author} {\bibfnamefont
  {S.}~\bibnamefont {Sun}}, \bibinfo {author} {\bibfnamefont {S.~H.}\
  \bibnamefont {Kang}}, \bibinfo {author} {\bibfnamefont {V.}~\bibnamefont
  {Tournat}},\ and\ \bibinfo {author} {\bibfnamefont {K.}~\bibnamefont
  {Bertoldi}},\ }\bibfield  {title} {\bibinfo {title} {Harnessing geometric
  frustration to form band gaps in acoustic channel lattices},\ }\href@noop {}
  {\bibfield  {journal} {\bibinfo  {journal} {Physical Review Letters}\
  }\textbf {\bibinfo {volume} {118}},\ \bibinfo {pages} {084302} (\bibinfo
  {year} {2017})}\BibitemShut {NoStop}%
\bibitem [{\citenamefont {Kang}\ \emph {et~al.}(2014)\citenamefont {Kang},
  \citenamefont {Shan}, \citenamefont {Ko{\v{s}}mrlj}, \citenamefont
  {Noorduin}, \citenamefont {Shian}, \citenamefont {Weaver}, \citenamefont
  {Clarke},\ and\ \citenamefont {Bertoldi}}]{kang2014complex}%
  \BibitemOpen
  \bibfield  {author} {\bibinfo {author} {\bibfnamefont {S.~H.}\ \bibnamefont
  {Kang}}, \bibinfo {author} {\bibfnamefont {S.}~\bibnamefont {Shan}}, \bibinfo
  {author} {\bibfnamefont {A.}~\bibnamefont {Ko{\v{s}}mrlj}}, \bibinfo {author}
  {\bibfnamefont {W.~L.}\ \bibnamefont {Noorduin}}, \bibinfo {author}
  {\bibfnamefont {S.}~\bibnamefont {Shian}}, \bibinfo {author} {\bibfnamefont
  {J.~C.}\ \bibnamefont {Weaver}}, \bibinfo {author} {\bibfnamefont {D.~R.}\
  \bibnamefont {Clarke}},\ and\ \bibinfo {author} {\bibfnamefont
  {K.}~\bibnamefont {Bertoldi}},\ }\bibfield  {title} {\bibinfo {title}
  {Complex ordered patterns in mechanical instability induced geometrically
  frustrated triangular cellular structures},\ }\href@noop {} {\bibfield
  {journal} {\bibinfo  {journal} {Physical Review Letters}\ }\textbf {\bibinfo
  {volume} {112}},\ \bibinfo {pages} {098701} (\bibinfo {year}
  {2014})}\BibitemShut {NoStop}%
\bibitem [{\citenamefont {Han}\ \emph {et~al.}(2008)\citenamefont {Han},
  \citenamefont {Shokef}, \citenamefont {Alsayed}, \citenamefont {Yunker},
  \citenamefont {Lubensky},\ and\ \citenamefont {Yodh}}]{han2008geometric}%
  \BibitemOpen
  \bibfield  {author} {\bibinfo {author} {\bibfnamefont {Y.}~\bibnamefont
  {Han}}, \bibinfo {author} {\bibfnamefont {Y.}~\bibnamefont {Shokef}},
  \bibinfo {author} {\bibfnamefont {A.~M.}\ \bibnamefont {Alsayed}}, \bibinfo
  {author} {\bibfnamefont {P.}~\bibnamefont {Yunker}}, \bibinfo {author}
  {\bibfnamefont {T.~C.}\ \bibnamefont {Lubensky}},\ and\ \bibinfo {author}
  {\bibfnamefont {A.~G.}\ \bibnamefont {Yodh}},\ }\bibfield  {title} {\bibinfo
  {title} {Geometric frustration in buckled colloidal monolayers},\ }\href@noop
  {} {\bibfield  {journal} {\bibinfo  {journal} {Nature}\ }\textbf {\bibinfo
  {volume} {456}},\ \bibinfo {pages} {898} (\bibinfo {year}
  {2008})}\BibitemShut {NoStop}%
\bibitem [{\citenamefont {Shokef}\ and\ \citenamefont
  {Lubensky}(2009)}]{ShokefLubensky2009}%
  \BibitemOpen
  \bibfield  {author} {\bibinfo {author} {\bibfnamefont {Y.}~\bibnamefont
  {Shokef}}\ and\ \bibinfo {author} {\bibfnamefont {T.~C.}\ \bibnamefont
  {Lubensky}},\ }\bibfield  {title} {\bibinfo {title} {Stripes, zigzags, and
  slow dynamics in buckled hard spheres},\ }\href@noop {} {\bibfield  {journal}
  {\bibinfo  {journal} {Physical Review Letters}\ }\textbf {\bibinfo {volume}
  {102}},\ \bibinfo {pages} {048303} (\bibinfo {year} {2009})}\BibitemShut
  {NoStop}%
\bibitem [{\citenamefont {Coulais}\ \emph {et~al.}(2016)\citenamefont
  {Coulais}, \citenamefont {Teomy}, \citenamefont {de~Reus}, \citenamefont
  {Shokef},\ and\ \citenamefont {van Hecke}}]{coulais2016}%
  \BibitemOpen
  \bibfield  {author} {\bibinfo {author} {\bibfnamefont {C.}~\bibnamefont
  {Coulais}}, \bibinfo {author} {\bibfnamefont {E.}~\bibnamefont {Teomy}},
  \bibinfo {author} {\bibfnamefont {K.}~\bibnamefont {de~Reus}}, \bibinfo
  {author} {\bibfnamefont {Y.}~\bibnamefont {Shokef}},\ and\ \bibinfo {author}
  {\bibfnamefont {M.}~\bibnamefont {van Hecke}},\ }\bibfield  {title} {\bibinfo
  {title} {Combinatorial design of textured mechanical metamaterials},\
  }\href@noop {} {\bibfield  {journal} {\bibinfo  {journal} {Nature}\ }\textbf
  {\bibinfo {volume} {535}},\ \bibinfo {pages} {529} (\bibinfo {year}
  {2016})}\BibitemShut {NoStop}%
\bibitem [{\citenamefont {Bertoldi}\ \emph {et~al.}(2017)\citenamefont
  {Bertoldi}, \citenamefont {Vitelli}, \citenamefont {Christensen},\ and\
  \citenamefont {van Hecke}}]{bertoldi2017flexible}%
  \BibitemOpen
  \bibfield  {author} {\bibinfo {author} {\bibfnamefont {K.}~\bibnamefont
  {Bertoldi}}, \bibinfo {author} {\bibfnamefont {V.}~\bibnamefont {Vitelli}},
  \bibinfo {author} {\bibfnamefont {J.}~\bibnamefont {Christensen}},\ and\
  \bibinfo {author} {\bibfnamefont {M.}~\bibnamefont {van Hecke}},\ }\bibfield
  {title} {\bibinfo {title} {Flexible mechanical metamaterials},\ }\href@noop
  {} {\bibfield  {journal} {\bibinfo  {journal} {Nature Reviews Materials}\
  }\textbf {\bibinfo {volume} {2}},\ \bibinfo {pages} {1} (\bibinfo {year}
  {2017})}\BibitemShut {NoStop}%
\bibitem [{\citenamefont {Meeussen}\ \emph
  {et~al.}(2020{\natexlab{a}})\citenamefont {Meeussen}, \citenamefont
  {O{\u{g}}uz}, \citenamefont {Shokef},\ and\ \citenamefont {van
  Hecke}}]{meeussen2020topological}%
  \BibitemOpen
  \bibfield  {author} {\bibinfo {author} {\bibfnamefont {A.~S.}\ \bibnamefont
  {Meeussen}}, \bibinfo {author} {\bibfnamefont {E.~C.}\ \bibnamefont
  {O{\u{g}}uz}}, \bibinfo {author} {\bibfnamefont {Y.}~\bibnamefont {Shokef}},\
  and\ \bibinfo {author} {\bibfnamefont {M.}~\bibnamefont {van Hecke}},\
  }\bibfield  {title} {\bibinfo {title} {Topological defects produce exotic
  mechanics in complex metamaterials},\ }\href@noop {} {\bibfield  {journal}
  {\bibinfo  {journal} {Nature Physics}\ }\textbf {\bibinfo {volume} {16}},\
  \bibinfo {pages} {307} (\bibinfo {year} {2020}{\natexlab{a}})}\BibitemShut
  {NoStop}%
\bibitem [{\citenamefont {Pisanty}\ \emph {et~al.}(2021)\citenamefont
  {Pisanty}, \citenamefont {O{\u{g}}uz}, \citenamefont {Nisoli},\ and\
  \citenamefont {Shokef}}]{pisanty2021putting}%
  \BibitemOpen
  \bibfield  {author} {\bibinfo {author} {\bibfnamefont {B.}~\bibnamefont
  {Pisanty}}, \bibinfo {author} {\bibfnamefont {E.~C.}\ \bibnamefont
  {O{\u{g}}uz}}, \bibinfo {author} {\bibfnamefont {C.}~\bibnamefont {Nisoli}},\
  and\ \bibinfo {author} {\bibfnamefont {Y.}~\bibnamefont {Shokef}},\
  }\bibfield  {title} {\bibinfo {title} {Putting a spin on metamaterials:
  Mechanical incompatibility as magnetic frustration},\ }\href@noop {}
  {\bibfield  {journal} {\bibinfo  {journal} {SciPost Physics}\ }\textbf
  {\bibinfo {volume} {10}},\ \bibinfo {pages} {136} (\bibinfo {year}
  {2021})}\BibitemShut {NoStop}%
\bibitem [{\citenamefont {Meeussen}\ \emph
  {et~al.}(2020{\natexlab{b}})\citenamefont {Meeussen}, \citenamefont
  {O{\u{g}}uz}, \citenamefont {van Hecke},\ and\ \citenamefont
  {Shokef}}]{meeussen2020response}%
  \BibitemOpen
  \bibfield  {author} {\bibinfo {author} {\bibfnamefont {A.~S.}\ \bibnamefont
  {Meeussen}}, \bibinfo {author} {\bibfnamefont {E.~C.}\ \bibnamefont
  {O{\u{g}}uz}}, \bibinfo {author} {\bibfnamefont {M.}~\bibnamefont {van
  Hecke}},\ and\ \bibinfo {author} {\bibfnamefont {Y.}~\bibnamefont {Shokef}},\
  }\bibfield  {title} {\bibinfo {title} {Response evolution of mechanical
  metamaterials under architectural transformations},\ }\href@noop {}
  {\bibfield  {journal} {\bibinfo  {journal} {New Journal of Physics}\ }\textbf
  {\bibinfo {volume} {22}},\ \bibinfo {pages} {023030} (\bibinfo {year}
  {2020}{\natexlab{b}})}\BibitemShut {NoStop}%
\bibitem [{\citenamefont {Deng}\ \emph {et~al.}(2020)\citenamefont {Deng},
  \citenamefont {Yu}, \citenamefont {Forte}, \citenamefont {Tournat},\ and\
  \citenamefont {Bertoldi}}]{deng2020characterization}%
  \BibitemOpen
  \bibfield  {author} {\bibinfo {author} {\bibfnamefont {B.}~\bibnamefont
  {Deng}}, \bibinfo {author} {\bibfnamefont {S.}~\bibnamefont {Yu}}, \bibinfo
  {author} {\bibfnamefont {A.~E.}\ \bibnamefont {Forte}}, \bibinfo {author}
  {\bibfnamefont {V.}~\bibnamefont {Tournat}},\ and\ \bibinfo {author}
  {\bibfnamefont {K.}~\bibnamefont {Bertoldi}},\ }\bibfield  {title} {\bibinfo
  {title} {Characterization, stability, and application of domain walls in
  flexible mechanical metamaterials},\ }\href@noop {} {\bibfield  {journal}
  {\bibinfo  {journal} {Proceedings of the National Academy of Sciences of the
  USA}\ }\textbf {\bibinfo {volume} {117}},\ \bibinfo {pages} {31002} (\bibinfo
  {year} {2020})}\BibitemShut {NoStop}%
\bibitem [{\citenamefont {Merrigan}\ \emph {et~al.}(2021)\citenamefont
  {Merrigan}, \citenamefont {Nisoli},\ and\ \citenamefont
  {Shokef}}]{merrigan2021topologically}%
  \BibitemOpen
  \bibfield  {author} {\bibinfo {author} {\bibfnamefont {C.}~\bibnamefont
  {Merrigan}}, \bibinfo {author} {\bibfnamefont {C.}~\bibnamefont {Nisoli}},\
  and\ \bibinfo {author} {\bibfnamefont {Y.}~\bibnamefont {Shokef}},\
  }\bibfield  {title} {\bibinfo {title} {Topologically protected steady cycles
  in an icelike mechanical metamaterial},\ }\href@noop {} {\bibfield  {journal}
  {\bibinfo  {journal} {Physical Review Research}\ }\textbf {\bibinfo {volume}
  {3}},\ \bibinfo {pages} {023174} (\bibinfo {year} {2021})}\BibitemShut
  {NoStop}%
\bibitem [{\citenamefont {Chen}\ \emph {et~al.}(2021)\citenamefont {Chen},
  \citenamefont {Pauly},\ and\ \citenamefont {Reis}}]{chen2021reprogrammable}%
  \BibitemOpen
  \bibfield  {author} {\bibinfo {author} {\bibfnamefont {T.}~\bibnamefont
  {Chen}}, \bibinfo {author} {\bibfnamefont {M.}~\bibnamefont {Pauly}},\ and\
  \bibinfo {author} {\bibfnamefont {P.~M.}\ \bibnamefont {Reis}},\ }\bibfield
  {title} {\bibinfo {title} {A reprogrammable mechanical metamaterial with
  stable memory},\ }\href@noop {} {\bibfield  {journal} {\bibinfo  {journal}
  {Nature}\ }\textbf {\bibinfo {volume} {589}},\ \bibinfo {pages} {386}
  (\bibinfo {year} {2021})}\BibitemShut {NoStop}%
\bibitem [{\citenamefont {Bense}\ and\ \citenamefont {van
  Hecke}(2021)}]{bense2021}%
  \BibitemOpen
  \bibfield  {author} {\bibinfo {author} {\bibfnamefont {H.}~\bibnamefont
  {Bense}}\ and\ \bibinfo {author} {\bibfnamefont {M.}~\bibnamefont {van
  Hecke}},\ }\bibfield  {title} {\bibinfo {title} {Complex pathways and memory
  in compressed corrugated sheets},\ }\href@noop {} {\bibfield  {journal}
  {\bibinfo  {journal} {Proceedings of the National Academy of Sciences of the
  USA}\ }\textbf {\bibinfo {volume} {118}},\ \bibinfo {pages} {e2111436118}
  (\bibinfo {year} {2021})}\BibitemShut {NoStop}%
\bibitem [{\citenamefont {Udani}\ and\ \citenamefont
  {Arrieta}(2022)}]{udani2021taming}%
  \BibitemOpen
  \bibfield  {author} {\bibinfo {author} {\bibfnamefont {J.~P.}\ \bibnamefont
  {Udani}}\ and\ \bibinfo {author} {\bibfnamefont {A.~F.}\ \bibnamefont
  {Arrieta}},\ }\bibfield  {title} {\bibinfo {title} {Taming geometric
  frustration by leveraging structural elasticity},\ }\href@noop {} {\bibfield
  {journal} {\bibinfo  {journal} {Materials and Design}\ }\textbf {\bibinfo
  {volume} {221}},\ \bibinfo {pages} {110809} (\bibinfo {year}
  {2022})}\BibitemShut {NoStop}%
\bibitem [{\citenamefont {Guo}\ \emph {et~al.}(2023)\citenamefont {Guo},
  \citenamefont {Guzman}, \citenamefont {Carpentier}, \citenamefont {Bartolo},\
  and\ \citenamefont {Coulais}}]{guo2021non}%
  \BibitemOpen
  \bibfield  {author} {\bibinfo {author} {\bibfnamefont {X.}~\bibnamefont
  {Guo}}, \bibinfo {author} {\bibfnamefont {M.}~\bibnamefont {Guzman}},
  \bibinfo {author} {\bibfnamefont {D.}~\bibnamefont {Carpentier}}, \bibinfo
  {author} {\bibfnamefont {D.}~\bibnamefont {Bartolo}},\ and\ \bibinfo {author}
  {\bibfnamefont {C.}~\bibnamefont {Coulais}},\ }\bibfield  {title} {\bibinfo
  {title} {Non-orientable order and non-commutative response in frustrated
  metamaterials},\ }\href@noop {} {\bibfield  {journal} {\bibinfo  {journal}
  {Nature}\ }\textbf {\bibinfo {volume} {618}},\ \bibinfo {pages} {506}
  (\bibinfo {year} {2023})}\BibitemShut {NoStop}%
\bibitem [{\citenamefont {Shohat}\ \emph {et~al.}(2022)\citenamefont {Shohat},
  \citenamefont {Hexner},\ and\ \citenamefont {Lahini}}]{shohat2021memory}%
  \BibitemOpen
  \bibfield  {author} {\bibinfo {author} {\bibfnamefont {D.}~\bibnamefont
  {Shohat}}, \bibinfo {author} {\bibfnamefont {D.}~\bibnamefont {Hexner}},\
  and\ \bibinfo {author} {\bibfnamefont {Y.}~\bibnamefont {Lahini}},\
  }\bibfield  {title} {\bibinfo {title} {Memory from coupled instabilities in
  crumpled sheets},\ }\href@noop {} {\bibfield  {journal} {\bibinfo  {journal}
  {Proceedings of the National Academy of Sciences of the USA}\ }\textbf
  {\bibinfo {volume} {119}},\ \bibinfo {pages} {e2200028119} (\bibinfo {year}
  {2022})}\BibitemShut {NoStop}%
\bibitem [{\citenamefont {Jules}\ \emph {et~al.}(2022)\citenamefont {Jules},
  \citenamefont {Reid}, \citenamefont {Daniels}, \citenamefont {Mungan},\ and\
  \citenamefont {Lechenault}}]{jules2022delicate}%
  \BibitemOpen
  \bibfield  {author} {\bibinfo {author} {\bibfnamefont {T.}~\bibnamefont
  {Jules}}, \bibinfo {author} {\bibfnamefont {A.}~\bibnamefont {Reid}},
  \bibinfo {author} {\bibfnamefont {K.~E.}\ \bibnamefont {Daniels}}, \bibinfo
  {author} {\bibfnamefont {M.}~\bibnamefont {Mungan}},\ and\ \bibinfo {author}
  {\bibfnamefont {F.}~\bibnamefont {Lechenault}},\ }\bibfield  {title}
  {\bibinfo {title} {Delicate memory structure of origami switches},\
  }\href@noop {} {\bibfield  {journal} {\bibinfo  {journal} {Physical Review
  Research}\ }\textbf {\bibinfo {volume} {4}},\ \bibinfo {pages} {013128}
  (\bibinfo {year} {2022})}\BibitemShut {NoStop}%
\bibitem [{\citenamefont {Lahini}\ \emph {et~al.}(2017)\citenamefont {Lahini},
  \citenamefont {Gottesman}, \citenamefont {Amir},\ and\ \citenamefont
  {Rubinstein}}]{lahini2017}%
  \BibitemOpen
  \bibfield  {author} {\bibinfo {author} {\bibfnamefont {Y.}~\bibnamefont
  {Lahini}}, \bibinfo {author} {\bibfnamefont {O.}~\bibnamefont {Gottesman}},
  \bibinfo {author} {\bibfnamefont {A.}~\bibnamefont {Amir}},\ and\ \bibinfo
  {author} {\bibfnamefont {S.~M.}\ \bibnamefont {Rubinstein}},\ }\bibfield
  {title} {\bibinfo {title} {Nonmonotonic aging and memory retention in
  disordered mechanical systems},\ }\href@noop {} {\bibfield  {journal}
  {\bibinfo  {journal} {Physical Review Letters}\ }\textbf {\bibinfo {volume}
  {118}},\ \bibinfo {pages} {085501} (\bibinfo {year} {2017})}\BibitemShut
  {NoStop}%
\bibitem [{\citenamefont {Shohat}\ and\ \citenamefont
  {Lahini}(2023)}]{shohat2023dissipation}%
  \BibitemOpen
  \bibfield  {author} {\bibinfo {author} {\bibfnamefont {D.}~\bibnamefont
  {Shohat}}\ and\ \bibinfo {author} {\bibfnamefont {Y.}~\bibnamefont
  {Lahini}},\ }\bibfield  {title} {\bibinfo {title} {Dissipation indicates
  memory formation in driven disordered systems},\ }\href@noop {} {\bibfield
  {journal} {\bibinfo  {journal} {Physical Review Letters}\ }\textbf {\bibinfo
  {volume} {130}},\ \bibinfo {pages} {048202} (\bibinfo {year}
  {2023})}\BibitemShut {NoStop}%
\bibitem [{\citenamefont {Yasuda}\ \emph {et~al.}(2021)\citenamefont {Yasuda},
  \citenamefont {Buskohl}, \citenamefont {Gillman}, \citenamefont {Murphey},
  \citenamefont {Stepney}, \citenamefont {Vaia},\ and\ \citenamefont
  {Raney}}]{Yasuda2021computing}%
  \BibitemOpen
  \bibfield  {author} {\bibinfo {author} {\bibfnamefont {H.}~\bibnamefont
  {Yasuda}}, \bibinfo {author} {\bibfnamefont {P.~R.}\ \bibnamefont {Buskohl}},
  \bibinfo {author} {\bibfnamefont {A.}~\bibnamefont {Gillman}}, \bibinfo
  {author} {\bibfnamefont {T.~D.}\ \bibnamefont {Murphey}}, \bibinfo {author}
  {\bibfnamefont {S.}~\bibnamefont {Stepney}}, \bibinfo {author} {\bibfnamefont
  {R.~A.}\ \bibnamefont {Vaia}},\ and\ \bibinfo {author} {\bibfnamefont
  {J.~R.}\ \bibnamefont {Raney}},\ }\bibfield  {title} {\bibinfo {title}
  {Mechanical computing},\ }\href@noop {} {\bibfield  {journal} {\bibinfo
  {journal} {Nature}\ }\textbf {\bibinfo {volume} {598}},\ \bibinfo {pages}
  {39} (\bibinfo {year} {2021})}\BibitemShut {NoStop}%
\bibitem [{\citenamefont {Treml}\ \emph {et~al.}(2018)\citenamefont {Treml},
  \citenamefont {Gillman}, \citenamefont {Buskohl},\ and\ \citenamefont
  {Vaia}}]{treml2018origami}%
  \BibitemOpen
  \bibfield  {author} {\bibinfo {author} {\bibfnamefont {B.}~\bibnamefont
  {Treml}}, \bibinfo {author} {\bibfnamefont {A.}~\bibnamefont {Gillman}},
  \bibinfo {author} {\bibfnamefont {P.}~\bibnamefont {Buskohl}},\ and\ \bibinfo
  {author} {\bibfnamefont {R.}~\bibnamefont {Vaia}},\ }\bibfield  {title}
  {\bibinfo {title} {Origami mechanologic},\ }\href@noop {} {\bibfield
  {journal} {\bibinfo  {journal} {Proceedings of the National Academy of
  Sciences of the USA}\ }\textbf {\bibinfo {volume} {115}},\ \bibinfo {pages}
  {6916} (\bibinfo {year} {2018})}\BibitemShut {NoStop}%
\bibitem [{\citenamefont {Mungan}\ \emph {et~al.}(2019)\citenamefont {Mungan},
  \citenamefont {Sastry}, \citenamefont {Dahmen},\ and\ \citenamefont
  {Regev}}]{mungan2019networks}%
  \BibitemOpen
  \bibfield  {author} {\bibinfo {author} {\bibfnamefont {M.}~\bibnamefont
  {Mungan}}, \bibinfo {author} {\bibfnamefont {S.}~\bibnamefont {Sastry}},
  \bibinfo {author} {\bibfnamefont {K.}~\bibnamefont {Dahmen}},\ and\ \bibinfo
  {author} {\bibfnamefont {I.}~\bibnamefont {Regev}},\ }\bibfield  {title}
  {\bibinfo {title} {Networks and hierarchies: How amorphous materials learn to
  remember},\ }\href@noop {} {\bibfield  {journal} {\bibinfo  {journal}
  {Physical Review Letters}\ }\textbf {\bibinfo {volume} {123}},\ \bibinfo
  {pages} {178002} (\bibinfo {year} {2019})}\BibitemShut {NoStop}%
\bibitem [{\citenamefont {Keim}\ \emph {et~al.}(2020)\citenamefont {Keim},
  \citenamefont {Hass}, \citenamefont {Kroger},\ and\ \citenamefont
  {Wieker}}]{keim2020global}%
  \BibitemOpen
  \bibfield  {author} {\bibinfo {author} {\bibfnamefont {N.~C.}\ \bibnamefont
  {Keim}}, \bibinfo {author} {\bibfnamefont {J.}~\bibnamefont {Hass}}, \bibinfo
  {author} {\bibfnamefont {B.}~\bibnamefont {Kroger}},\ and\ \bibinfo {author}
  {\bibfnamefont {D.}~\bibnamefont {Wieker}},\ }\bibfield  {title} {\bibinfo
  {title} {Global memory from local hysteresis in an amorphous solid},\
  }\href@noop {} {\bibfield  {journal} {\bibinfo  {journal} {Physical Review
  Research}\ }\textbf {\bibinfo {volume} {2}},\ \bibinfo {pages} {012004}
  (\bibinfo {year} {2020})}\BibitemShut {NoStop}%
\bibitem [{\citenamefont {Shokef}\ \emph {et~al.}(2011)\citenamefont {Shokef},
  \citenamefont {Souslov},\ and\ \citenamefont {Lubensky}}]{shokef-PNAS-2011}%
  \BibitemOpen
  \bibfield  {author} {\bibinfo {author} {\bibfnamefont {Y.}~\bibnamefont
  {Shokef}}, \bibinfo {author} {\bibfnamefont {A.}~\bibnamefont {Souslov}},\
  and\ \bibinfo {author} {\bibfnamefont {T.~C.}\ \bibnamefont {Lubensky}},\
  }\bibfield  {title} {\bibinfo {title} {Order-by-disorder in the
  antiferromagnetic ising model on an elastic triangular lattice},\ }\href@noop
  {} {\bibfield  {journal} {\bibinfo  {journal} {Proceedings of the National
  Academy of Sciences of the USA}\ }\textbf {\bibinfo {volume} {108}},\
  \bibinfo {pages} {11804} (\bibinfo {year} {2011})}\BibitemShut {NoStop}%
\bibitem [{\citenamefont {Morrison}\ \emph {et~al.}(2013)\citenamefont
  {Morrison}, \citenamefont {Nelson},\ and\ \citenamefont
  {Nisoli}}]{morrison2013unhappy}%
  \BibitemOpen
  \bibfield  {author} {\bibinfo {author} {\bibfnamefont {M.~J.}\ \bibnamefont
  {Morrison}}, \bibinfo {author} {\bibfnamefont {T.~R.}\ \bibnamefont
  {Nelson}},\ and\ \bibinfo {author} {\bibfnamefont {C.}~\bibnamefont
  {Nisoli}},\ }\bibfield  {title} {\bibinfo {title} {Unhappy vertices in
  artificial spin ice: new degeneracies from vertex frustration},\ }\href@noop
  {} {\bibfield  {journal} {\bibinfo  {journal} {New Journal of Physics}\
  }\textbf {\bibinfo {volume} {15}},\ \bibinfo {pages} {045009} (\bibinfo
  {year} {2013})}\BibitemShut {NoStop}%
\bibitem [{\citenamefont {Wang}\ \emph {et~al.}(2006)\citenamefont {Wang},
  \citenamefont {Nisoli}, \citenamefont {Freitas}, \citenamefont {Li},
  \citenamefont {McConville}, \citenamefont {Cooley}, \citenamefont {Lund},
  \citenamefont {Samarth}, \citenamefont {Leighton}, \citenamefont {Crespi},\
  and\ \citenamefont {P.}}]{wang2006ASI}%
  \BibitemOpen
  \bibfield  {author} {\bibinfo {author} {\bibfnamefont {R.~F.}\ \bibnamefont
  {Wang}}, \bibinfo {author} {\bibfnamefont {C.}~\bibnamefont {Nisoli}},
  \bibinfo {author} {\bibfnamefont {R.~S.}\ \bibnamefont {Freitas}}, \bibinfo
  {author} {\bibfnamefont {J.}~\bibnamefont {Li}}, \bibinfo {author}
  {\bibfnamefont {W.}~\bibnamefont {McConville}}, \bibinfo {author}
  {\bibfnamefont {B.~J.}\ \bibnamefont {Cooley}}, \bibinfo {author}
  {\bibfnamefont {M.~S.}\ \bibnamefont {Lund}}, \bibinfo {author}
  {\bibfnamefont {N.}~\bibnamefont {Samarth}}, \bibinfo {author} {\bibfnamefont
  {C.}~\bibnamefont {Leighton}}, \bibinfo {author} {\bibfnamefont {V.~H.}\
  \bibnamefont {Crespi}},\ and\ \bibinfo {author} {\bibfnamefont
  {S.}~\bibnamefont {P.}},\ }\bibfield  {title} {\bibinfo {title} {Artificial
  ‘spin ice’ in a geometrically frustrated lattice of nanoscale
  ferromagnetic islands},\ }\href@noop {} {\bibfield  {journal} {\bibinfo
  {journal} {Nature}\ }\textbf {\bibinfo {volume} {439}},\ \bibinfo {pages}
  {303} (\bibinfo {year} {2006})}\BibitemShut {NoStop}%
\bibitem [{\citenamefont {Ding}\ and\ \citenamefont {van
  Hecke}(2022)}]{ding2022sequential}%
  \BibitemOpen
  \bibfield  {author} {\bibinfo {author} {\bibfnamefont {J.}~\bibnamefont
  {Ding}}\ and\ \bibinfo {author} {\bibfnamefont {M.}~\bibnamefont {van
  Hecke}},\ }\bibfield  {title} {\bibinfo {title} {Sequential snapping and
  pathways in a mechanical metamaterial},\ }\href@noop {} {\bibfield  {journal}
  {\bibinfo  {journal} {The Journal of Chemical Physics}\ }\textbf {\bibinfo
  {volume} {156}},\ \bibinfo {pages} {204902} (\bibinfo {year}
  {2022})}\BibitemShut {NoStop}%
\bibitem [{\citenamefont {Nisoli}\ \emph {et~al.}(2013)\citenamefont {Nisoli},
  \citenamefont {Moessner},\ and\ \citenamefont
  {Schiffer}}]{nisoli2013colloquium}%
  \BibitemOpen
  \bibfield  {author} {\bibinfo {author} {\bibfnamefont {C.}~\bibnamefont
  {Nisoli}}, \bibinfo {author} {\bibfnamefont {R.}~\bibnamefont {Moessner}},\
  and\ \bibinfo {author} {\bibfnamefont {P.}~\bibnamefont {Schiffer}},\
  }\bibfield  {title} {\bibinfo {title} {Colloquium: Artificial spin ice:
  Designing and imaging magnetic frustration},\ }\href@noop {} {\bibfield
  {journal} {\bibinfo  {journal} {Reviews of Modern Physics}\ }\textbf
  {\bibinfo {volume} {85}},\ \bibinfo {pages} {1473} (\bibinfo {year}
  {2013})}\BibitemShut {NoStop}%
\bibitem [{\citenamefont {Skj{\ae}rv{\o}}\ \emph {et~al.}(2020)\citenamefont
  {Skj{\ae}rv{\o}}, \citenamefont {Marrows}, \citenamefont {Stamps},\ and\
  \citenamefont {Heyderman}}]{skjaervo2020advances}%
  \BibitemOpen
  \bibfield  {author} {\bibinfo {author} {\bibfnamefont {S.~H.}\ \bibnamefont
  {Skj{\ae}rv{\o}}}, \bibinfo {author} {\bibfnamefont {C.~H.}\ \bibnamefont
  {Marrows}}, \bibinfo {author} {\bibfnamefont {R.~L.}\ \bibnamefont
  {Stamps}},\ and\ \bibinfo {author} {\bibfnamefont {L.~J.}\ \bibnamefont
  {Heyderman}},\ }\bibfield  {title} {\bibinfo {title} {Advances in artificial
  spin ice},\ }\href@noop {} {\bibfield  {journal} {\bibinfo  {journal} {Nature
  Reviews Physics}\ }\textbf {\bibinfo {volume} {2}},\ \bibinfo {pages} {13}
  (\bibinfo {year} {2020})}\BibitemShut {NoStop}%
\bibitem [{\citenamefont {Gilbert}\ \emph {et~al.}(2016)\citenamefont
  {Gilbert}, \citenamefont {Nisoli},\ and\ \citenamefont
  {Schiffer}}]{gilbert2016frustration}%
  \BibitemOpen
  \bibfield  {author} {\bibinfo {author} {\bibfnamefont {I.}~\bibnamefont
  {Gilbert}}, \bibinfo {author} {\bibfnamefont {C.}~\bibnamefont {Nisoli}},\
  and\ \bibinfo {author} {\bibfnamefont {P.}~\bibnamefont {Schiffer}},\
  }\bibfield  {title} {\bibinfo {title} {Frustration by design},\ }\href@noop
  {} {\bibfield  {journal} {\bibinfo  {journal} {Physics Today}\ }\textbf
  {\bibinfo {volume} {69}},\ \bibinfo {pages} {54} (\bibinfo {year}
  {2016})}\BibitemShut {NoStop}%
\bibitem [{\citenamefont {Nisoli}\ \emph {et~al.}(2017)\citenamefont {Nisoli},
  \citenamefont {Kapaklis},\ and\ \citenamefont
  {Schiffer}}]{nisoli2017deliberate}%
  \BibitemOpen
  \bibfield  {author} {\bibinfo {author} {\bibfnamefont {C.}~\bibnamefont
  {Nisoli}}, \bibinfo {author} {\bibfnamefont {V.}~\bibnamefont {Kapaklis}},\
  and\ \bibinfo {author} {\bibfnamefont {P.}~\bibnamefont {Schiffer}},\
  }\bibfield  {title} {\bibinfo {title} {Deliberate exotic magnetism via
  frustration and topology},\ }\href@noop {} {\bibfield  {journal} {\bibinfo
  {journal} {Nature Physics}\ }\textbf {\bibinfo {volume} {13}},\ \bibinfo
  {pages} {200} (\bibinfo {year} {2017})}\BibitemShut {NoStop}%
\bibitem [{\citenamefont {Stamps}(2014)}]{stamps2014artificial}%
  \BibitemOpen
  \bibfield  {author} {\bibinfo {author} {\bibfnamefont {R.~L.}\ \bibnamefont
  {Stamps}},\ }\bibfield  {title} {\bibinfo {title} {Artificial spin ice: The
  unhappy wanderer},\ }\href@noop {} {\bibfield  {journal} {\bibinfo  {journal}
  {Nature Physics}\ }\textbf {\bibinfo {volume} {10}},\ \bibinfo {pages} {623}
  (\bibinfo {year} {2014})}\BibitemShut {NoStop}%
\bibitem [{\citenamefont {Gilbert}\ \emph {et~al.}(2014)\citenamefont
  {Gilbert}, \citenamefont {Chern}, \citenamefont {Zhang}, \citenamefont
  {O'Brien}, \citenamefont {Fore}, \citenamefont {Nisoli},\ and\ \citenamefont
  {Schiffer}}]{gilbert2014emergent}%
  \BibitemOpen
  \bibfield  {author} {\bibinfo {author} {\bibfnamefont {I.}~\bibnamefont
  {Gilbert}}, \bibinfo {author} {\bibfnamefont {G.-W.}\ \bibnamefont {Chern}},
  \bibinfo {author} {\bibfnamefont {S.}~\bibnamefont {Zhang}}, \bibinfo
  {author} {\bibfnamefont {L.}~\bibnamefont {O'Brien}}, \bibinfo {author}
  {\bibfnamefont {B.}~\bibnamefont {Fore}}, \bibinfo {author} {\bibfnamefont
  {C.}~\bibnamefont {Nisoli}},\ and\ \bibinfo {author} {\bibfnamefont
  {P.}~\bibnamefont {Schiffer}},\ }\bibfield  {title} {\bibinfo {title}
  {Emergent ice rule and magnetic charge screening from vertex frustration in
  artificial spin ice},\ }\href@noop {} {\bibfield  {journal} {\bibinfo
  {journal} {Nature Physics}\ }\textbf {\bibinfo {volume} {10}},\ \bibinfo
  {pages} {670} (\bibinfo {year} {2014})}\BibitemShut {NoStop}%
\bibitem [{\citenamefont {Saglam}\ \emph {et~al.}(2022)\citenamefont {Saglam},
  \citenamefont {Duzgun}, \citenamefont {Kargioti}, \citenamefont {Harle},
  \citenamefont {Zhang}, \citenamefont {Bingham}, \citenamefont {Lao},
  \citenamefont {Gilbert}, \citenamefont {Sklenar}, \citenamefont {Watts} \emph
  {et~al.}}]{saglam2022entropy}%
  \BibitemOpen
  \bibfield  {author} {\bibinfo {author} {\bibfnamefont {H.}~\bibnamefont
  {Saglam}}, \bibinfo {author} {\bibfnamefont {A.}~\bibnamefont {Duzgun}},
  \bibinfo {author} {\bibfnamefont {A.}~\bibnamefont {Kargioti}}, \bibinfo
  {author} {\bibfnamefont {N.}~\bibnamefont {Harle}}, \bibinfo {author}
  {\bibfnamefont {X.}~\bibnamefont {Zhang}}, \bibinfo {author} {\bibfnamefont
  {N.~S.}\ \bibnamefont {Bingham}}, \bibinfo {author} {\bibfnamefont
  {Y.}~\bibnamefont {Lao}}, \bibinfo {author} {\bibfnamefont {I.}~\bibnamefont
  {Gilbert}}, \bibinfo {author} {\bibfnamefont {J.}~\bibnamefont {Sklenar}},
  \bibinfo {author} {\bibfnamefont {J.~D.}\ \bibnamefont {Watts}}, \emph
  {et~al.},\ }\bibfield  {title} {\bibinfo {title} {Entropy-driven order in an
  array of nanomagnets},\ }\href@noop {} {\bibfield  {journal} {\bibinfo
  {journal} {Nature Physics}\ }\textbf {\bibinfo {volume} {18}},\ \bibinfo
  {pages} {706} (\bibinfo {year} {2022})}\BibitemShut {NoStop}%
\bibitem [{\citenamefont {Rodr{\'\i}guez-Gallo}\ \emph
  {et~al.}(2023)\citenamefont {Rodr{\'\i}guez-Gallo}, \citenamefont
  {Ortiz-Ambriz}, \citenamefont {Nisoli},\ and\ \citenamefont
  {Tierno}}]{rodriguez2023geometrical}%
  \BibitemOpen
  \bibfield  {author} {\bibinfo {author} {\bibfnamefont {C.}~\bibnamefont
  {Rodr{\'\i}guez-Gallo}}, \bibinfo {author} {\bibfnamefont {A.}~\bibnamefont
  {Ortiz-Ambriz}}, \bibinfo {author} {\bibfnamefont {C.}~\bibnamefont
  {Nisoli}},\ and\ \bibinfo {author} {\bibfnamefont {P.}~\bibnamefont
  {Tierno}},\ }\bibfield  {title} {\bibinfo {title} {Geometrical control of
  topological charge transfer in shakti-cairo colloidal ice},\ }\href@noop {}
  {\bibfield  {journal} {\bibinfo  {journal} {Communications Physics}\ }\textbf
  {\bibinfo {volume} {6}},\ \bibinfo {pages} {113} (\bibinfo {year}
  {2023})}\BibitemShut {NoStop}%
\bibitem [{\citenamefont {Chern}\ \emph {et~al.}(2013)\citenamefont {Chern},
  \citenamefont {Morrison},\ and\ \citenamefont
  {Nisoli}}]{chern2013degeneracy}%
  \BibitemOpen
  \bibfield  {author} {\bibinfo {author} {\bibfnamefont {G.-W.}\ \bibnamefont
  {Chern}}, \bibinfo {author} {\bibfnamefont {M.~J.}\ \bibnamefont
  {Morrison}},\ and\ \bibinfo {author} {\bibfnamefont {C.}~\bibnamefont
  {Nisoli}},\ }\bibfield  {title} {\bibinfo {title} {Degeneracy and criticality
  from emergent frustration in artificial spin ice},\ }\href@noop {} {\bibfield
   {journal} {\bibinfo  {journal} {Physical Review Letters}\ }\textbf {\bibinfo
  {volume} {111}},\ \bibinfo {pages} {177201} (\bibinfo {year}
  {2013})}\BibitemShut {NoStop}%
\bibitem [{\citenamefont {Lao}\ \emph {et~al.}(2018)\citenamefont {Lao},
  \citenamefont {Caravelli}, \citenamefont {Sheikh}, \citenamefont {Sklenar},
  \citenamefont {Gardeazabal}, \citenamefont {Watts}, \citenamefont {Albrecht},
  \citenamefont {Scholl}, \citenamefont {Dahmen}, \citenamefont {Nisoli},\ and\
  \citenamefont {Schiffer}}]{lao2018classical}%
  \BibitemOpen
  \bibfield  {author} {\bibinfo {author} {\bibfnamefont {Y.}~\bibnamefont
  {Lao}}, \bibinfo {author} {\bibfnamefont {F.}~\bibnamefont {Caravelli}},
  \bibinfo {author} {\bibfnamefont {M.}~\bibnamefont {Sheikh}}, \bibinfo
  {author} {\bibfnamefont {J.}~\bibnamefont {Sklenar}}, \bibinfo {author}
  {\bibfnamefont {D.}~\bibnamefont {Gardeazabal}}, \bibinfo {author}
  {\bibfnamefont {J.~D.}\ \bibnamefont {Watts}}, \bibinfo {author}
  {\bibfnamefont {A.~M.}\ \bibnamefont {Albrecht}}, \bibinfo {author}
  {\bibfnamefont {A.}~\bibnamefont {Scholl}}, \bibinfo {author} {\bibfnamefont
  {K.}~\bibnamefont {Dahmen}}, \bibinfo {author} {\bibfnamefont
  {C.}~\bibnamefont {Nisoli}},\ and\ \bibinfo {author} {\bibfnamefont
  {P.}~\bibnamefont {Schiffer}},\ }\bibfield  {title} {\bibinfo {title}
  {Classical topological order in the kinetics of artificial spin ice},\
  }\href@noop {} {\bibfield  {journal} {\bibinfo  {journal} {Nature Physics}\
  }\textbf {\bibinfo {volume} {14}},\ \bibinfo {pages} {723} (\bibinfo {year}
  {2018})}\BibitemShut {NoStop}%
\bibitem [{\citenamefont {Keim}\ \emph {et~al.}(2019)\citenamefont {Keim},
  \citenamefont {Paulsen}, \citenamefont {Zeravcic}, \citenamefont {Sastry},\
  and\ \citenamefont {Nagel}}]{keim2019memory}%
  \BibitemOpen
  \bibfield  {author} {\bibinfo {author} {\bibfnamefont {N.~C.}\ \bibnamefont
  {Keim}}, \bibinfo {author} {\bibfnamefont {J.~D.}\ \bibnamefont {Paulsen}},
  \bibinfo {author} {\bibfnamefont {Z.}~\bibnamefont {Zeravcic}}, \bibinfo
  {author} {\bibfnamefont {S.}~\bibnamefont {Sastry}},\ and\ \bibinfo {author}
  {\bibfnamefont {S.~R.}\ \bibnamefont {Nagel}},\ }\bibfield  {title} {\bibinfo
  {title} {Memory formation in matter},\ }\href@noop {} {\bibfield  {journal}
  {\bibinfo  {journal} {Reviews of Modern Physics}\ }\textbf {\bibinfo {volume}
  {91}},\ \bibinfo {pages} {035002} (\bibinfo {year} {2019})}\BibitemShut
  {NoStop}%
\bibitem [{\citenamefont {van Hecke}(2021)}]{van2021profusion}%
  \BibitemOpen
  \bibfield  {author} {\bibinfo {author} {\bibfnamefont {M.}~\bibnamefont {van
  Hecke}},\ }\bibfield  {title} {\bibinfo {title} {Profusion of transition
  pathways for interacting hysterons},\ }\href@noop {} {\bibfield  {journal}
  {\bibinfo  {journal} {Physical Review E}\ }\textbf {\bibinfo {volume}
  {104}},\ \bibinfo {pages} {054608} (\bibinfo {year} {2021})}\BibitemShut
  {NoStop}%
\bibitem [{\citenamefont {Singh}\ \emph {et~al.}(2023)\citenamefont {Singh},
  \citenamefont {Teunisse}, \citenamefont {Labousse},\ and\ \citenamefont {van
  Hecke}}]{singh2023emergent}%
  \BibitemOpen
  \bibfield  {author} {\bibinfo {author} {\bibfnamefont {A.}~\bibnamefont
  {Singh}}, \bibinfo {author} {\bibfnamefont {M.}~\bibnamefont {Teunisse}},
  \bibinfo {author} {\bibfnamefont {M.}~\bibnamefont {Labousse}},\ and\
  \bibinfo {author} {\bibfnamefont {M.}~\bibnamefont {van Hecke}},\ }\bibfield
  {title} {\bibinfo {title} {Emergent computing in a non-abelian
  metamaterial},\ }\href@noop {} {\bibfield  {journal} {\bibinfo  {journal}
  {Bulletin of the American Physical Society}\ } (\bibinfo {year}
  {2023})}\BibitemShut {NoStop}%
\bibitem [{sup()}]{supplemental}%
  \BibitemOpen
  \href@noop {} {}\bibinfo {howpublished} {See Supplemental Material at
  \url{https://youtu.be/9FKM54Lb5gU} for movie of non-Abelian response in
  experiments of the Chaco metamaterial undergoing the same operations at
  different sequences.}\BibitemShut {Stop}%
\bibitem [{\citenamefont {Fruchart}\ \emph {et~al.}(2020)\citenamefont
  {Fruchart}, \citenamefont {Zhou},\ and\ \citenamefont
  {Vitelli}}]{Fruchart_2020_dualities}%
  \BibitemOpen
  \bibfield  {author} {\bibinfo {author} {\bibfnamefont {M.}~\bibnamefont
  {Fruchart}}, \bibinfo {author} {\bibfnamefont {Y.}~\bibnamefont {Zhou}},\
  and\ \bibinfo {author} {\bibfnamefont {V.}~\bibnamefont {Vitelli}},\
  }\bibfield  {title} {\bibinfo {title} {Dualities and non-{A}belian
  mechanics},\ }\href@noop {} {\bibfield  {journal} {\bibinfo  {journal}
  {Nature}\ }\textbf {\bibinfo {volume} {577}},\ \bibinfo {pages} {636}
  (\bibinfo {year} {2020})}\BibitemShut {NoStop}%
\bibitem [{\citenamefont {Liu}\ \emph {et~al.}(2023)\citenamefont {Liu},
  \citenamefont {Domino}, \citenamefont {de~Dinechin}, \citenamefont
  {Taffetani},\ and\ \citenamefont {Vella}}]{liu2021frustrating}%
  \BibitemOpen
  \bibfield  {author} {\bibinfo {author} {\bibfnamefont {M.}~\bibnamefont
  {Liu}}, \bibinfo {author} {\bibfnamefont {L.}~\bibnamefont {Domino}},
  \bibinfo {author} {\bibfnamefont {I.~D.}\ \bibnamefont {de~Dinechin}},
  \bibinfo {author} {\bibfnamefont {M.}~\bibnamefont {Taffetani}},\ and\
  \bibinfo {author} {\bibfnamefont {D.}~\bibnamefont {Vella}},\ }\bibfield
  {title} {\bibinfo {title} {Snap-induced morphing: From a single bistable
  shell to the origin of shape bifurcation in interacting shells},\ }\href@noop
  {} {\bibfield  {journal} {\bibinfo  {journal} {Journal of the Mechanics and
  Physics of Solids}\ }\textbf {\bibinfo {volume} {170}},\ \bibinfo {pages}
  {105116} (\bibinfo {year} {2023})}\BibitemShut {NoStop}%
\bibitem [{\citenamefont {Hexner}(2021)}]{hexner2021training}%
  \BibitemOpen
  \bibfield  {author} {\bibinfo {author} {\bibfnamefont {D.}~\bibnamefont
  {Hexner}},\ }\bibfield  {title} {\bibinfo {title} {Training nonlinear elastic
  functions: nonmonotonic, sequence dependent and bifurcating},\ }\href@noop {}
  {\bibfield  {journal} {\bibinfo  {journal} {Soft Matter}\ }\textbf {\bibinfo
  {volume} {17}},\ \bibinfo {pages} {4407} (\bibinfo {year}
  {2021})}\BibitemShut {NoStop}%
\bibitem [{\citenamefont {Coulais}\ \emph {et~al.}(2018)\citenamefont
  {Coulais}, \citenamefont {Sabbadini}, \citenamefont {Vink},\ and\
  \citenamefont {van Hecke}}]{coulais2018multi}%
  \BibitemOpen
  \bibfield  {author} {\bibinfo {author} {\bibfnamefont {C.}~\bibnamefont
  {Coulais}}, \bibinfo {author} {\bibfnamefont {A.}~\bibnamefont {Sabbadini}},
  \bibinfo {author} {\bibfnamefont {F.}~\bibnamefont {Vink}},\ and\ \bibinfo
  {author} {\bibfnamefont {M.}~\bibnamefont {van Hecke}},\ }\bibfield  {title}
  {\bibinfo {title} {Multi-step self-guided pathways for shape-changing
  metamaterials},\ }\href@noop {} {\bibfield  {journal} {\bibinfo  {journal}
  {Nature}\ }\textbf {\bibinfo {volume} {561}},\ \bibinfo {pages} {512}
  (\bibinfo {year} {2018})}\BibitemShut {NoStop}%
\bibitem [{\citenamefont {Yuan}\ \emph {et~al.}(2021)\citenamefont {Yuan},
  \citenamefont {Chen}, \citenamefont {Yao}, \citenamefont {Guo}, \citenamefont
  {Huang}, \citenamefont {Peng}, \citenamefont {Xu}, \citenamefont {Lv},
  \citenamefont {Tao}, \citenamefont {Duan}, \citenamefont {Liao},
  \citenamefont {Yao}, \citenamefont {Li}, \citenamefont {Lei}, \citenamefont
  {Chen}, \citenamefont {Hong},\ and\ \citenamefont {Fang}}]{yuan2021recent}%
  \BibitemOpen
  \bibfield  {author} {\bibinfo {author} {\bibfnamefont {X.}~\bibnamefont
  {Yuan}}, \bibinfo {author} {\bibfnamefont {M.}~\bibnamefont {Chen}}, \bibinfo
  {author} {\bibfnamefont {Y.}~\bibnamefont {Yao}}, \bibinfo {author}
  {\bibfnamefont {X.}~\bibnamefont {Guo}}, \bibinfo {author} {\bibfnamefont
  {Y.}~\bibnamefont {Huang}}, \bibinfo {author} {\bibfnamefont
  {Z.}~\bibnamefont {Peng}}, \bibinfo {author} {\bibfnamefont {B.}~\bibnamefont
  {Xu}}, \bibinfo {author} {\bibfnamefont {B.}~\bibnamefont {Lv}}, \bibinfo
  {author} {\bibfnamefont {R.}~\bibnamefont {Tao}}, \bibinfo {author}
  {\bibfnamefont {S.}~\bibnamefont {Duan}}, \bibinfo {author} {\bibfnamefont
  {H.}~\bibnamefont {Liao}}, \bibinfo {author} {\bibfnamefont {K.}~\bibnamefont
  {Yao}}, \bibinfo {author} {\bibfnamefont {Y.}~\bibnamefont {Li}}, \bibinfo
  {author} {\bibfnamefont {H.}~\bibnamefont {Lei}}, \bibinfo {author}
  {\bibfnamefont {X.}~\bibnamefont {Chen}}, \bibinfo {author} {\bibfnamefont
  {G.}~\bibnamefont {Hong}},\ and\ \bibinfo {author} {\bibfnamefont
  {D.}~\bibnamefont {Fang}},\ }\bibfield  {title} {\bibinfo {title} {Recent
  progress in the design and fabrication of multifunctional structures based on
  metamaterials},\ }\href@noop {} {\bibfield  {journal} {\bibinfo  {journal}
  {Current Opinion in Solid State and Materials Science}\ }\textbf {\bibinfo
  {volume} {25}},\ \bibinfo {pages} {100883} (\bibinfo {year}
  {2021})}\BibitemShut {NoStop}%
\bibitem [{\citenamefont {Bossart}\ \emph {et~al.}(2021)\citenamefont
  {Bossart}, \citenamefont {Dykstra}, \citenamefont {Van~der Laan},\ and\
  \citenamefont {Coulais}}]{bossart2021oligomodal}%
  \BibitemOpen
  \bibfield  {author} {\bibinfo {author} {\bibfnamefont {A.}~\bibnamefont
  {Bossart}}, \bibinfo {author} {\bibfnamefont {D.~M.}\ \bibnamefont
  {Dykstra}}, \bibinfo {author} {\bibfnamefont {J.}~\bibnamefont {Van~der
  Laan}},\ and\ \bibinfo {author} {\bibfnamefont {C.}~\bibnamefont {Coulais}},\
  }\bibfield  {title} {\bibinfo {title} {Oligomodal metamaterials with
  multifunctional mechanics},\ }\href@noop {} {\bibfield  {journal} {\bibinfo
  {journal} {Proceedings of the National Academy of Sciences of the USA}\
  }\textbf {\bibinfo {volume} {118}},\ \bibinfo {pages} {e2018610118} (\bibinfo
  {year} {2021})}\BibitemShut {NoStop}%
\bibitem [{\citenamefont {Lindeman}\ and\ \citenamefont
  {Nagel}(2021)}]{lindeman2021multiple}%
  \BibitemOpen
  \bibfield  {author} {\bibinfo {author} {\bibfnamefont {C.~W.}\ \bibnamefont
  {Lindeman}}\ and\ \bibinfo {author} {\bibfnamefont {S.~R.}\ \bibnamefont
  {Nagel}},\ }\bibfield  {title} {\bibinfo {title} {Multiple memory formation
  in glassy landscapes},\ }\href@noop {} {\bibfield  {journal} {\bibinfo
  {journal} {Science Advances}\ }\textbf {\bibinfo {volume} {7}},\ \bibinfo
  {pages} {eabg7133} (\bibinfo {year} {2021})}\BibitemShut {NoStop}%
\bibitem [{\citenamefont {Keim}\ and\ \citenamefont
  {Paulsen}(2021)}]{keim2021multiperiodic}%
  \BibitemOpen
  \bibfield  {author} {\bibinfo {author} {\bibfnamefont {N.~C.}\ \bibnamefont
  {Keim}}\ and\ \bibinfo {author} {\bibfnamefont {J.~D.}\ \bibnamefont
  {Paulsen}},\ }\bibfield  {title} {\bibinfo {title} {Multiperiodic orbits from
  interacting soft spots in cyclically sheared amorphous solids},\ }\href@noop
  {} {\bibfield  {journal} {\bibinfo  {journal} {Science Advances}\ }\textbf
  {\bibinfo {volume} {7}},\ \bibinfo {pages} {eabg7685} (\bibinfo {year}
  {2021})}\BibitemShut {NoStop}%
\bibitem [{\citenamefont {Szulc}\ \emph {et~al.}(2022)\citenamefont {Szulc},
  \citenamefont {Mungan},\ and\ \citenamefont {Regev}}]{Szulc_JCP_2022}%
  \BibitemOpen
  \bibfield  {author} {\bibinfo {author} {\bibfnamefont {A.}~\bibnamefont
  {Szulc}}, \bibinfo {author} {\bibfnamefont {M.}~\bibnamefont {Mungan}},\ and\
  \bibinfo {author} {\bibfnamefont {I.}~\bibnamefont {Regev}},\ }\bibfield
  {title} {\bibinfo {title} {Cooperative effects driving the multi-periodic
  dynamics of cyclically sheared amorphous solids},\ }\href@noop {} {\bibfield
  {journal} {\bibinfo  {journal} {Journal of Chemical Physics}\ }\textbf
  {\bibinfo {volume} {156}},\ \bibinfo {pages} {164506} (\bibinfo {year}
  {2022})}\BibitemShut {NoStop}%
\bibitem [{\citenamefont {Kwakernaak}\ and\ \citenamefont {van
  Hecke}(2023)}]{kwakernaak2023counting}%
  \BibitemOpen
  \bibfield  {author} {\bibinfo {author} {\bibfnamefont {L.~J.}\ \bibnamefont
  {Kwakernaak}}\ and\ \bibinfo {author} {\bibfnamefont {M.}~\bibnamefont {van
  Hecke}},\ }\bibfield  {title} {\bibinfo {title} {Counting and sequential
  information processing in mechanical metamaterials},\ }\href@noop {}
  {\bibfield  {journal} {\bibinfo  {journal} {arXiv preprint arXiv:2302.06947}\
  } (\bibinfo {year} {2023})}\BibitemShut {NoStop}%
\bibitem [{\citenamefont {Lamberty}\ \emph {et~al.}(2013)\citenamefont
  {Lamberty}, \citenamefont {Papanikolaou},\ and\ \citenamefont
  {Henley}}]{lamberty2013classical}%
  \BibitemOpen
  \bibfield  {author} {\bibinfo {author} {\bibfnamefont {R.~Z.}\ \bibnamefont
  {Lamberty}}, \bibinfo {author} {\bibfnamefont {S.}~\bibnamefont
  {Papanikolaou}},\ and\ \bibinfo {author} {\bibfnamefont {C.~L.}\ \bibnamefont
  {Henley}},\ }\bibfield  {title} {\bibinfo {title} {Classical topological
  order in {A}belian and non-{A}belian generalized height models},\ }\href@noop
  {} {\bibfield  {journal} {\bibinfo  {journal} {Physical Review Letters}\
  }\textbf {\bibinfo {volume} {111}},\ \bibinfo {pages} {245701} (\bibinfo
  {year} {2013})}\BibitemShut {NoStop}%
\end{thebibliography}%

\end{document}